\newcommand{\CLASSINPUTtoptextmargin}{21mm}
\documentclass[journal,draftclsnofoot,onecolumn,12pt,twoside]{IEEEtran}
\usepackage[T1]{fontenc}
\usepackage{color}
\definecolor{mycolor}{RGB}{0, 0, 150}
\usepackage{algorithm} 
\usepackage{algpseudocode}
\usepackage[neverdecrease]{paralist}
\usepackage{graphicx}
\usepackage{amssymb}
\usepackage{amsfonts}
\usepackage[cmex10]{amsmath}
\usepackage{dsfont}
\usepackage{float}
\usepackage{url}
\usepackage{bm}
\usepackage{hhline}
\usepackage{enumerate}
\usepackage{pgf,tikz}
\usepackage{mathrsfs}
\usetikzlibrary{arrows}
\usepackage{array}
\usepackage{tabulary}
\newcolumntype{K}[1]{>{\centering\arraybackslash}p{#1}}

 \usepackage[colorlinks=true]{hyperref}

\newtheorem{theorem}{Theorem}[section]

\newtheorem{definition}[theorem]{Definition}

\usepackage[cmintegrals]{newtxmath}

\ifCLASSINFOpdf
\else
\fi

\begin{document}
%
\title{A Lattice Based Joint Encryption, Encoding and Modulation Scheme}

%
%
%
%

\author{Khadijeh Bagheri,  Taraneh Eghlidos, Mohammad-Reza Sadeghi,  Daniel Panario ~\IEEEmembership{Senior Member,~IEEE,}
\thanks{K. Bagheri  and M.-R Sadeghi are with Faculty
of Mathematics and Computer Science, Amirkabir University of Technology (Tehran Polytechnic), Tehran, Iran (emails: kbagheri@aut.ac.ir and msadeghi@aut.ac.ir).

T. Eghlidos is with Electronics Research Institute, Sharif University of Technology, Iran
(e-mail: teghlidos@sharif.edu).

D. Panario is with School of Mathematics and Statistics, Carleton University, Canada
(e-mail: daniel@math.carleton.ca).

Part of this work has been presented in \cite{Mina}.
}}
\maketitle
\begin{abstract}
A new nonlinear Rao-Nam like symmetric key encryption scheme is presented in this paper.  QC-LDPC lattices  that are practically implementable in high dimensions due to their low complexity encoding and decoding algorithms, are used in our design.
Then, a joint scheme is proposed which is capable of encrypting, encoding and data modulation simultaneously.
The proposed cryptosystem withstands all variants of chosen plaintext attacks applied on Rao-Nam like cryptosystems due to its nonlinearity. The sparseness of the parity-check matrix of QC-LDPC lattices, quasi-cyclic nature of their generator and parity-check matrices, simple hardware structure for generating intentional error vector, permutation and nonlinear functions, result in a small key size for our scheme. The lattice codes related to the lattices used in this paper have high rate which are suitable for bandlimited AWGN channels.
Therefore, the joint scheme based on these lattices facilitates secure, reliable and efficient data transmission in bandlimited AWGN channels.

\end{abstract}

\IEEEpeerreviewmaketitle

\IEEEdisplaynontitleabstractindextext

%
\IEEEpeerreviewmaketitle

\section{Introduction}\label{sec:introduction}
\IEEEPARstart{T}{he} main objective in large scale and high speed communication networks is the design of a reliable and secure data transmission system.
In conventional secure communication systems, channel encoding is used at the physical layer for error correction while encryption  is done at a higher layer to make the communication confidential. Nowadays, since many communication devices are constrained in resources or are becoming portable, it is needed to enhance security without increasing computational or hardware complexity.
Rao was the first that joined encryption and error correction (based on a linear block code) in a single step \cite{Rao}
to resolve the issue in an insecure and unreliable channel efficiently.
The main purpose of  this  scheme is to provide both security and reliable data transmission at the same time using a symmetric-key cryptosystem.

Rao defines the proposed cryptosystem based on binary Goppa codes. It is proved that this cryptosystem is not secure against chosen-plaintext attacks. To overcome this weakness, Rao and Nam have introduced a revised symmetric-key cryptosystem \cite{Rao-Nam}, \cite{Rao-Nam2}. In this paper we use RN to abbreviate Rao and Nam cryptosystem.  Struik and Tilburg  have generalized the RN scheme to any finite field $\mathbb{F}_q$ \cite{STRUIK}. Barbero and Ytrehus  have reduced the  secret key size of the RN scheme, while maintaining the security level of the system \cite{Barbero-Ytrehus}.
The previous proposed RN-like schemes  have very large key size or low information rate  that makes them impractical.
Several schemes have been presented in the literature to modify  RN, reduce the key size, increase information rate  and improve security against known attacks. Some of them use quasi-cyclic (QC) codes to reduce the key size and use low-density parity-check (LDPC) codes to increase the information rate \cite{Eghlidos}, \cite{Hooshmand-ISC}.
Unlike the previous RN-like schemes that scramble, permute or change the codeword bits,  the proposed joint schemes based on QC-LDPC codes in  \cite{Dakhilalian}, \cite{Esmaeili1} and \cite{Esmaeili2}, randomly puncture the codeword corresponding to a plaintext.
The security and the key sizes of these schemes have been significantly improved.
Another joint scheme has been proposed based on low-density lattice codes (LDLCs)  \cite{LDLC} that has practical decoder \cite{Hooshmand}.

All of these RN-like schemes are  vulnerable to the message-resent attack \cite{Berson}.
It is an interesting idea to consider nonlinear layers for  RN-like schemes to provide their resistance against well known chosen plaintext attacks.   The RN-like schemes, named SECC (secret error-correcting codes),  use nonlinear codes, or block chaining technique to withstand against the chosen plaintext attacks, while retaining the error correction performance \cite{SECC}.
ECBC (Error Correction Based Cipher) is another  RN-like scheme that   uses a nonlinear function $f$ in its structure \cite{ECBC}.
In parallel, a differential-style attack (chosen-plaintext attack) against  ECBC and SECC has been proposed in  \cite{Differential-Cryptanalysis}. It is shown that  the non-linear function $f$  used in ECBC  is particularly vulnerable to this attack and its secret key  is recovered in constant time.
The security of \cite{SECC} and \cite{ECBC} is enhanced to resist the differential attack proposed in \cite{Differential-Cryptanalysis} by considering a special design for the nonlinear function $f$ \cite{Deepthi-Nonlinear}.
The authors claim that the proposed scheme is the most efficient secure channel coding scheme among  the previous code based schemes \cite{Deepthi-Nonlinear}.

There is another approach for joining encryption and  encoding in a single step. The proposed schemes in this approach use error correcting codes as a replacement of the AES (Advanced Encryption Standard) diffusion layer \cite{HD}, \cite{GLOBECOM}.
The message-resend attack   \cite{Berson} is not  effective on these joint AES-coding schemes.
In spite of providing error correction, small key size (equal to AES) and data security against linear and differential attacks, these schemes suffer from higher computational complexities compared to RN-like schemes.

In the RN-like schemes, the  security is embedded in the channel code to enhance the security of the entire system. They are fast and easy from an implementation standpoint.
The earlier RN-like schemes have smaller error correction capability than error correction codes.
Indeed,  to achieve significant error correction capacity, the schemes' parameters
 have to be large which lead to large key size and high computational complexity. However, this issue has been resolved in  recent RN-like schemes \cite{Deepthi-Nonlinear}.
Low power consumption for RN-like schemes is an important advantage that leads us to concentrate on RN-like schemes. This allows us to use the same hardware components for error correction and security.

In the conventional joint schemes, modulation is applied on the resulting point that is sent to the channel. In this paper, we propose an alternative secure wireless communication model that uses a coded-modulation scheme based on lattices to design an RN-like scheme
to merge encryption, channel encoding and modulation in a single step.
Using lattices in designing RN-like schemes is rare due to the high complexity of lattice encoding and decoding even for legitimate users.
We use a special type of lattices that provide proper efficiency for the proposed scheme.

In general, a digital modulator is used at the output of the channel encoder, which serves as an interface to the communication channels. The modulator produces signal waveforms that transmit information.
When treating a channel coding as a separate operation independent of modulation, the coded sequences generally have a smaller channel symbol duration than the uncoded sequences for the same information rate. Accordingly, the power spectral density (PSD) of the channel signals changes essentially. On the other hand, if the modulation is designed in conjunction with the channel coding, error correction can be achieved without leading to any essential changes in the PSD. The advantages of the coded-modulation schemes, that are the schemes in which coding and modulation are combined, are highlighted here. This importance is featured more when designing coding techniques for high-SNR (signal to noise ratio) regimes or bandlimited channels \cite{Ungerboeck}.
Therefore, our goal is to use a proper coded-modulation scheme in our cryptosystem to provide a secure communication for high-SNR or bandlimited channels.

Based on Shannon's capacity theorem, an optimal block code for a bandlimited additive white Gaussian noise (AWGN) channel  consists of a dense packing of the code points within a sphere in a high-dimensional Euclidean space.
Lattices and trellis codes are two main categories of packings, for moderate coding gains at moderate complexity \cite{Forney}.
Lattices are most of the densest known packings. Construction-A of lattices \cite{conway} are not among the densest packings but define a natural mapping from codewords of the underlying code to lattice points in the Euclidean space. Thus, they are coded-modulation schemes.

A lattice code is defined by a lattice $\Lambda$ and a shaping region $B\subset \mathbb{R}^n$, where its codewords are all the lattice points that belong to the shaping region $B$.
Lattice codes are a suitable coded-modulation scheme on bandlimited (high-SNR) AWGN channels \cite{Forney}.
In this paper, we use a lattice code of some kind of construction-A lattices.

LDPC lattices are the first family of modern lattices \cite{sadeghi1, sadeghi2} with practical decoding algorithms in high dimensions \cite{Safarnejad}. It has been shown that these lattices can achieve desirable performance when compared to other modern lattices \cite{Hassan}. A special sub-class of LDPC lattices, QC-LDPC lattices, is a construction-A lattices that have practical encoder as well  \cite{QC-LDPC-Lattices}. Due to the easy construction, linear encoding and decoding complexity and good error performance of the  QC-LDPC lattices, their lattice codes have the potential to become a practical and efficient coding scheme for the AWGN channel. Furthermore, QC-LDPC lattices have many applications in the wireless communications \cite{Full-Duplex}, \cite{Hassan-ISIT},  as well as in designing  symmetric and asymmetric key encryption schemes \cite{Mina}, \cite{Mina-Public}.

In this paper, we  exploit the characteristics of  QC-LDPC lattice codes and the efficiency of this coded-modulations to merge encryption, channel encoding and modulation in a single step.
Due to the hardware reuse, such scheme improves the transmission  efficiency, reduces the system processing delay and saves in hardware usage.
More flexibility in terms of design and technology used for the construction is given, as well. The sparse nature of the parity-check matrix of the  QC-LDPC lattices can be exploited to reduce the key size. Moreover, good error performance and linear encoding and decoding (time and space) complexities (in terms of the lattice dimension) of these lattice codes provide an efficient joint scheme  that is suitable for secure communication in bandlimited (high-SNR) AWGN channels.



This paper is organized as follows.
In Section  \ref{sec:preliminaries}, we present some lattice notation used in this paper.
In Section \ref{sec:Encryption},  a symmetric key encryption scheme based on  QC-LDPC lattices is presented.
We describe a joint encryption, encoding and modulation scheme  based on  these lattices in Section \ref{sec:joint encryption}.
In Section \ref{sec:Key Size and Complexity}, we give details about the key sizes, computational complexity of  encryption and decryption.
Section \ref{sec:attacks} is devoted to the security analysis and numerical results related to the security of our scheme against different attacks. The comparison with other RN-like schemes and a summary of the paper are given in Section \ref{sec:Comparison} and Section \ref{sec:Conclusion}, respectively. 
\section{Preliminaries}~\label{sec:preliminaries}
\subsection{Construction of Lattices from Codes}
Let $\mathbb{R}^{m}$ be the $m$-dimensional real vector space with Euclidean norm. An $n$-dimensional  \emph{lattice} is a discrete additive subgroup of $\mathbb{R}^m$ that is presented by the set of all integer linear combinations of a given basis of $n$ linearly independent vectors in  $\mathbb{R}^{m}$ \cite{conway}.
Let $\mathbf{b_1},\mathbf{b_2},\dots,\mathbf{b_n}\in \mathbb{R}^m$, be the basis vectors of the  lattice $\Lambda$. The \emph{generator matrix} for the lattice  is the matrix $\mathbf{B}$,  where  $\mathbf{b}_i$ is its $i$-th row. Therefore, the lattice $\Lambda$ is defined as $\Lambda=\{\boldsymbol{\xi}\cdot \mathbf{B} | ~\boldsymbol{\xi} \in \mathbb{Z}^{n}\}.$
If $m=n$,  the lattice is  \emph{full rank}.
For more details about lattices and their properties, see \cite{conway}.

There are many ways to construct lattices from linear codes; many properties of such lattices can be related to the properties of their underlying  codes \cite{conway}. In our work Construction-A of lattices is used that are explained in the sequel.

Let $\mathcal{C}=C \left[n, k \right]\subseteq\mathbb{Z}_{p}^n$ be a linear code with dimension $k$ and length $n$, where $p$ is a prime number. A lattice $\Lambda$ constructed from the code $\mathcal{C}$ based on Construction-A is defined by
$$\Lambda=p\mathbb{Z}^n+\phi\left(\mathcal{C}\right)=\{p\mathbf{z}+\phi(\mathbf{c}): \mathbf{c}\in \mathcal{C}, \mathbf{z}\in \mathbb{Z}^n\},$$
where $\phi\colon\mathbb{Z}_p^n\rightarrow\mathbb{R}^n$ is the embedding function which maps a vector in $\mathbb{Z}_p^n$ to its real version \cite{conway}.
 Indeed, Construction-A is a method for generating a lattice by ``lifting'' a linear code $\mathcal{C}$  to the Euclidean space.  In this paper, we use binary linear  codes and lattices with $p=2$.
We consider a subclass of LDPC lattices \cite{sadeghi2} that can be obtained from Construction-A \cite{sadeghi3}.
It is interesting to focus on the quasi-cyclic (QC) version of these lattices for cryptographic purposes.
\begin{definition}
Let $\mathcal{C}=C \left[n, k \right]$ be a binary linear QC-LDPC code. A \emph{QC-LDPC lattice} $\Lambda$ is a lattice based on Construction-A along with a binary linear QC-LDPC code $\mathcal{C}$ which is defined as
$\Lambda=\{\mathbf{x}\in \mathbb{Z}^n ~|~ \mathbf{H}_{qc}\cdot{\mathbf x}^T\equiv 0 \;\;(\bmod\,2)\}$,
where $\mathbf{H}_{qc}$ is the parity-check matrix of $\mathcal{C}$ \cite{QC-LDPC-Lattices}.
\end{definition}

The generator matrix of a QC-LDPC lattice $\Lambda$ is the following ${n\times n}$ matrix
\begin{equation}~\label{generator matrix}
\small
\mathbf{G_{\Lambda}}=\left[\begin{array}{cc}
\mathbf{I_{k}}& \mathbf{A_{k\times (n-k)}} \\
\mathbf{0_{(n-k)\times k}} & 2\mathbf{I_{n-k}}
\end{array}\right]=\left[\begin{array}{ll}
\quad \quad \quad\mathbf{G_{\mathcal{C}}} \\
\mathbf{0_{(n-k)\times k}} & 2\mathbf{I_{n-k}}
\end{array}\right],
\end{equation}
where $\mathbf{0}$ stands for the zero block, $\mathbf{I}$ for an identity block and
$\mathbf{G_{\mathcal{C}}}=\left[\begin{array}{cc}
\mathbf{I_{k}}& \mathbf{A_{k\times (n-k)}}
\end{array}\right]_{k\times n}$
is the systematic form of the generator matrix of  $\mathcal{C}$ \cite{Hassan}.
\subsubsection{Encoding of QC-LDPC lattices}\label{LDPC lattice encoding}
Based on a method for encoding Construction-A lattices in \cite{conway},  the encoding  of  LDPC lattices is introduced
in \cite{Hassan} as follows.
In order to find a decoding method for these family of lattices, a translated sublattice of lattice $\Lambda$, generated by (\ref{generator matrix}), is considered. First, the codewords' components of the binary code $\mathcal{C}$ are converted into $\pm 1$, that is,  $0$ is converted to $-1$ and $1$ is converted to $1$.
Then, the set $ \Lambda(\mathcal{C})=\{ \mathbf{c}+4\mathbf{z} \, |\, \mathbf{c}\in \mathcal{C},\,\, \mathbf{z}\in \mathbb{Z}^n\}$, where  $\mathcal{C}$ is a  QC-LDPC code with $\pm1$ components, is a lattice which is closed under the  addition $\bm{\lambda}_1\oplus\bm{\lambda}_2\triangleq\bm{\lambda}_1+\bm{\lambda}_2+(1,\ldots ,1)$, for all $\bm{\lambda}_1,\bm{\lambda}_2\in\Lambda(\mathcal{C})$.
Therefore, the encoding algorithm of sublattice $\Lambda(\mathcal{C})$ for any integer row vector $\boldsymbol{\xi}\in \mathbb{Z}^n$ can be done with the generator matrix (\ref{generator matrix})  as $E(\boldsymbol{\xi})= 2 \boldsymbol{\xi}\cdot \mathbf{G_{\Lambda}}-\mathbf{1}$,
where $\mathbf{1}=(1,\ldots ,1)$, $E$ is the encoding function and $E(\boldsymbol{\xi})$ is a point of the lattice $\Lambda(\mathcal{C})$ \cite{Hassan}.
Let the points of the lattice $\Lambda(\mathcal{C})$ be transmitted over an unconstrained AWGN channel with noise variance $\sigma^2$. Then, its volume-to-noise ratio  is defined as $\mbox {VNR}=\frac{4^{(2n-k)/n}}{2\pi e \sigma^2}$.
\subsubsection{Decoding of QC-LDPC lattices}
Let $\mathbf{x}=\mathbf{c}+4\mathbf{z}$ be the transmitted lattice vector of $ \Lambda(\mathcal{C})$ and  $\mathbf{y}=\mathbf{c}+4\mathbf{z}+\mathbf{n}$ be the received  vector from the AWGN channel,
where $\mathbf{n}\sim \mathcal{N}(0,\sigma^2)$.
A soft decision message passing algorithm, that is sum-product algorithm (SPA), is presented in \cite{QC-LDPC-Lattices} for decoding QC-LDPC lattices  that has low implementation complexity and memory requirements.
These lattices can be decoded in linear time complexity in terms of dimension.
\subsection{RDF-QC-LDPC codes}
To have efficient decoding and good performance for QC-LDPC codes, their parity-check matrix should be free of length-$4$ cycles.

In this paper, we consider  QC-LDPC codes with  rate $R=(n_0- 1)/n_0$  with  parity-check matrix described by
 \begin{equation}\label{parity check}
    \mathbf{H}_{qc}=\left[
                       \begin{array}{c|c|c|c}
                         \mathbf{H}_0  & \mathbf{H}_1  & \cdots  & \mathbf{H}_{n_0-1}
                       \end{array}
                     \right],
    \end{equation}
where $\textbf{H}_0, \ldots, \textbf{H}_{n_0-1}$ are $b\times b$ circulant matrices and have low row/column Hamming weight.

Some algebraic approaches for  code design are possible with this particular form of the parity-check matrix. However, some of them impose constraints on the code length.
The proposed techniques based on \emph{Difference Families} (DFs) such as \emph{Extended Difference Families} (EDFs) and \emph{Random Difference Families} (RDFs), loosen the constraints using computer aided procedures \cite{Baldi-book}.

Using these techniques and given the number of circulant blocks $n_0$  with column weight $d_v$, the size of the circulant matrices, $b$, is chosen to ensure the absence of length-$4$ cycles in the associated code \cite{Baldi-book}.

For $d_v=5$ and $n_0=8$, the size of circulant matrices is $b=187$ by the RDF construction technique \cite{Baldi-book}. Therefore, the RDF-QC-LDPC code with these  parameters ($d_v$ and $n_0$)
 has code length $1496$, dimension $1309$, rate $R = 7/8$ and  row weight $d_c = 40$ \cite{Baldi-book}. However, using the same parameters, the size of the circulant matrix for the EDF-based solution is  $b > 200$.
The minimum length of QC-LDPC code based on EDF technique for $d_v=5$ and $n_0=8$ is $ 1600$ \cite{Baldi-book}. Hence, the RDF-based approach allows to design shorter codes compared to those designed through the EDF approach.

\begin{figure}[h]
\centerline{\includegraphics[width=3.5in]{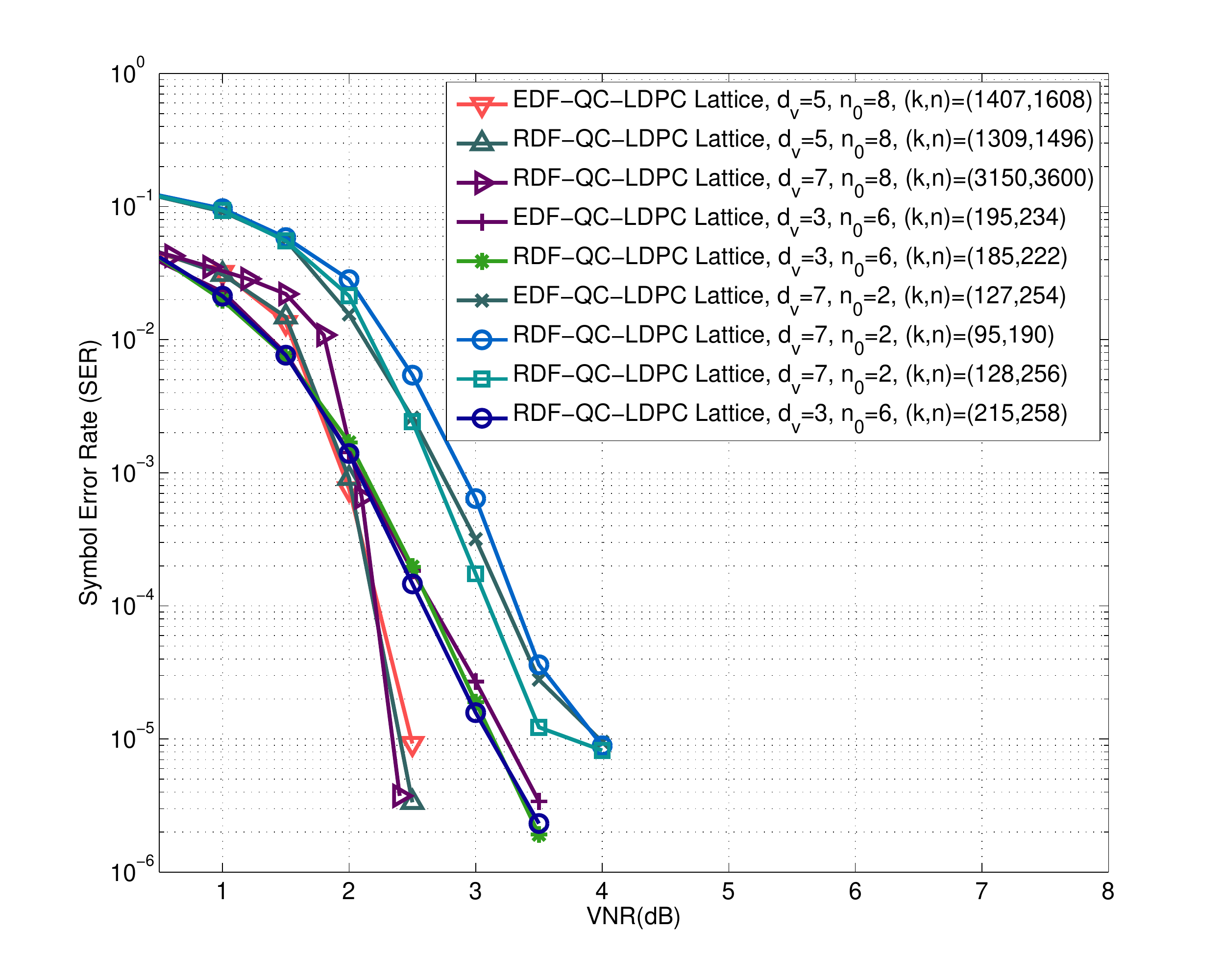}}
\caption{Error performance of EDF and RDF based QC-LDPC lattices.}\label{RDF.EDF-simulation}
\end{figure}

%
Simulation results of symbol error rate (SER) versus VNR  of  QC-LDPC lattices based on EDF and RDF over the AWGN channel are presented in   \figurename~\ref{RDF.EDF-simulation}.  According to the results of this figure, RDF based QC-LDPC lattices outperform EDF based QC-LDPC lattices in smaller dimensions. For example, in  SER$=10^{-5}$, an RDF based QC-LDPC lattice with parameters $(k,n)=(1309,1496)$, $d_v=5$ and $n_0=8$ has $0.05$dB better performance compared to  an EDF based QC-LDPC lattice with $(k,n)=(1407,1608)$, $d_v=5$ and $n_0=8$.

The results of \figurename~\ref{RDF.EDF-simulation} also show that by using underlying codes with higher rates and  increasing the $n_0$ value, which results in a decrease of $d_v$ value, we gain a considerable performance improvement.
For example, in the SER of $10^{-5}$, using an RDF based QC-LDPC lattice with $(k,n)=(215,258)$, $d_v=3$, $n_0=6$ gives us $0.5$dB performance improvement compared to the same dimension RDF based QC-LDPC lattice  with $(k,n)=(128,256)$, $d_v=7$, $n_0=2$.

Interesting aspects of the design technique based on RDFs,  which is important
for cryptographic applications, are generating a large number of equivalent codes with the same code length, dimension and column weight, as well as designing shorter codes compared to those designed based on DFs and EDFs  approaches.
All these good features and the easy construction  of RDF-QC-LDPC codes lead us to  use  them as the underlying codes used in our lattices.

In the rest of this paper, whenever we use a QC-LDPC lattice, we mean that the lattice is constructed using an RDF QC-LDPC code as its underlying code.
\section{Symmetric key cryptosystem based on QC-LDPC lattices}~\label{sec:Encryption}
In this work, we design a new symmetric key encryption scheme  using a special type of lattices, namely  QC-LDPC lattices.
Defining an appropriate nonlinear function, we enhance the security of the cryptosystem against chosen plaintext attack, differential or linear type attacks. We adapt the method presented in \cite{Function-f} to design an invertible nonlinear mapping $F$ for our encryption scheme where all the operations are performed over the field of real numbers $\mathbb{R}$.

\begin{figure}[h]
	\begin{center}
\definecolor{qqqqff}{rgb}{0.,0.,1.}
\begin{tikzpicture}[line cap=round,line join=round,>=triangle 45,x=0.9cm,y=0.9cm]
\clip(-5.91331923190833,-1.499142517555464) rectangle (5.161336712460304,5.198171753387659);
\draw (-2.36,3.4)-- (-1.08,3.4);
\draw (-2.36,3.4)-- (-2.36,2.26);
\draw (-1.08,3.4)-- (-1.08,2.26);
\draw (-2.36,2.26)-- (-1.08,2.26);
\draw (-1.08,2.26)-- (-2.36,2.26);
\draw (-1.74,4.02)-- (3.38,4.02);
\draw [->] (-1.74,4.02) -- (-1.74,3.4);
\draw (-1.72,2.26)-- (-1.72,1.8);
\draw (-1.72,1.8)-- (-0.9,1.8);
\draw [->] (0.9,4.66) -- (0.9,4.02);
\draw (-0.22,3.4)-- (1.06,3.4);
\draw (-0.22,3.4)-- (-0.22,2.26);
\draw (1.06,3.4)-- (1.06,2.26);
\draw (-0.22,2.26)-- (1.06,2.26);
\draw (1.06,2.26)-- (-0.22,2.26);
\draw [->] (0.42,4.02) -- (0.42,3.4);
\draw [->] (0.44,2.26) -- (0.43273,1.24);
\draw [->] (-0.9,1.8) -- (-0.8763636363636375,1.238181818181818);
\draw (-1.69455,1.24)-- (3.4,1.24);
\draw (-1.69455,1.24)-- (-1.694545454545456,0.47454545454545544);
\draw (-1.694545454545456,0.47454545454545544)-- (3.4,0.47);
\draw [->] (0.9018681224191829,0.4722288828315322) -- (0.9,-0.3364);
\draw [->] (-5.1127272727272745,0.8381818181818186) -- (-1.6945476139056432,0.8381817102009543);
\draw (-1.5706906030789811,5.1721423537455275) node[anchor=north west] {Input vector $\mathbf{a}$ ($n$-tuple)};
\draw (-5.41260123786735745,1.393416069659905) node[anchor=north west] {Control vector $\mathbf{b}$};
\draw (-1.2451582767810327,-0.46615985416359645) node[anchor=north west] {Output vector $F(\mathbf{a},\mathbf{b})$};
\draw (1.4527153927969545,2.828523576088912) node[anchor=north west] {$...$};
\draw (1.4527153927969545,1.8785228323964707) node[anchor=north west] {$...$};
\draw (-2.003670291701091,3.131715302799265) node[anchor=north west] {$F_0$};
\draw (0.058033449929322176,3.091289739237885) node[anchor=north west] {$F_1$};
\draw (0.28037404951691575,1.2285220887040296) node[anchor=north west] {MUX};
\draw (-4.6026078701825297,0.867883743361959) node[anchor=north west] { ($d$-tuple)};
\draw (2.744254148228917,3.44)-- (4.024254148228916,3.44);
\draw (2.744254148228917,3.44)-- (2.744254148228917,2.3);
\draw (4.024254148228916,3.44)-- (4.024254148228916,2.3);
\draw (2.744254148228917,2.3)-- (4.024254148228916,2.3);
\draw (4.024254148228916,2.3)-- (2.744254148228917,2.3);
\draw [->] (3.3842541482289166,4.02) -- (3.3842541482289157,3.44);
\draw (3.404254148228917,2.3)-- (3.404254148228917,1.8);
\draw (3.404254148228917,1.8)-- (2.5042541482289167,1.8);
\draw [->] (2.5042541482289167,1.8) -- (2.5188041482289165,1.24);
\draw (2.5188041482289165,1.24)-- (3.404254148228917,1.23818);
\draw (3.404254148228917,1.23818)-- (3.4097041482289168,0.47);
\draw (2.8473973356645867,3.192353648141336) node[anchor=north west] {$F_{2^d -1}$};
\draw (0.061225176639677,-0.9512666169001621) node[anchor=north west] { ($n$-tuple)};
\end{tikzpicture}
 \caption{The structure of nonlinear mapping $F$.}
\label{function}
\end{center}
\end{figure}
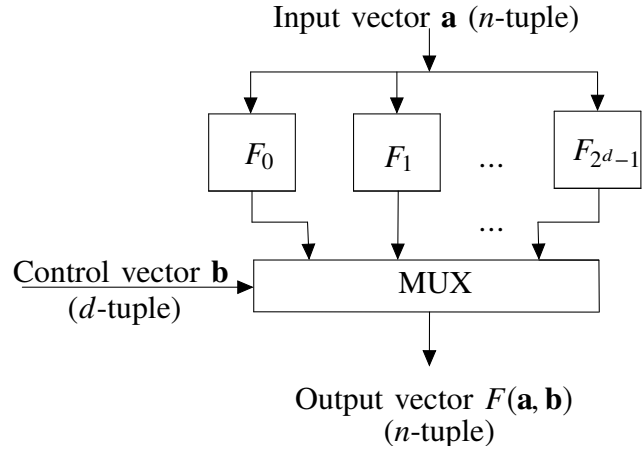

In the structure of this mapping, we use some $n\times n$ linear transformations represented by an $n\times n$ binary matrix.  We construct the transformation matrices using a companion matrix of a primitive polynomial  in  $\mathbb{F}_2[x]$ described as follows. The linear transformations corresponding to these invertible matrices are represented by the linear operators  $F_0, \ldots, ,F_{2^{d}-1}$, for an integer number $d> 0$. The input vector of the mapping  $F$ is $\mathbf{z}=(\mathbf{a}, \mathbf{b})$, where $\mathbf{a}$ is an $n$-tuple vector and $\mathbf{b}$ is a $d$-tuple binary vector.
The input to each linear transformation is an $n$-tuple vector $\mathbf{a}$ in which $a_i \in \mathbb{Z}$, for $i=0, \ldots, n-1$. Each linear operator $F_j$, for $j=0, \ldots, 2^{d}-1$, transforms an $n$-tuple input to an $n$-tuple output. Then,
the output of the linear transformations are passed
through a multiplexer, controlled by the $d$-tuple vector $\mathbf{b}$
which serves as the control line. Indeed, the control line of the multiplexer is used to select one of the outputs of the linear
transformations.
The overall structure of the mapping is shown  in \figurename~\ref{function}.

In this paper, we construct the corresponding matrices for the linear transformations used in the architecture of $F$ based on the following mathematical method. In this architecture, the applied linear transformations need to have maximal period. Indeed, the corresponding matrix $\mathbf{U}$ should have maximal order, that is, $\mathbf{U}^{2^n -1}=\mathbf{I}$, where $\mathbf{I}$ is the $n\times n$ identity matrix, and $\mathbf{U}^{m}\neq\mathbf{I}$, for $m< 2^n -1$.

Let $g\in \mathbb{F}_q[x]$ be a nonzero polynomial of degree $n\geq 1$. If $g(0) \neq 0$, the order of $g$ is the least
positive integer $e$, $1\leq e \leq q^n -1$, such that $g(x)| x^e -1$ and it is denoted by $ord(g)$ \cite{Daniel-Book}. It is known that the order of a primitive polynomial $g$ of degree $n$ is equal to $q^n -1$.
Let $g(x)=x^n+a_{n-1}x^{n-1}+\cdots +a_{1}x+a_0$, where $a_0\neq 0$, Then, $C(g)$ is  the
companion matrix of  $g$ as
\begin{equation*}\label{c(f)}
\small
C(g)=\left[ {\begin{array}{ccccc}
   0 & 1 & 0 &  \cdots    & 0  \\
   0 & 0 & 1 & \cdots  & 0  \\
    \vdots  &  \vdots  &  \vdots  &  \ddots   & 0  \\
   0 & 0 & 0 &  \cdots   & 1  \\
    - {a_0} & - {a_1} & - {a_2} &  \cdots   &  - {a_{n-1}}  \\
\end{array}} \right]_{n\times n}.
\end{equation*}
Since $\det(C(g)) = (-1)^n a_0$ and $g(0)=a_0\neq 0$,  the companion matrix  of $g$ is an invertible matrix, that is, $C(g)\in GL_n(q)$ \cite{Darafsheh}. Furthermore, the order of the polynomial $g$ is related to the order of companion matrix \cite{Darafsheh}. It can be verified that the order of $C(g)$ as an element of $GL_n(q)$ is equal to the order of  $g$.
It can be shown that the inverse of a companion matrix of the polynomial $g(x)=x^n+a_{n-1}x^{n-1}+a_{n-2}x^{n-2}+\cdots +a_{1}x+a_0$ is a companion matrix of the polynomial $r(x)=x^n+\frac{a_{1}}{a_0}x^{n-1}+\frac{a_{2}}{a_0}x^{n-2}+\cdots +\frac{a_{n-1}}{a_0}x+\frac{1}{a_0}$.

Let us consider a primitive  polynomial $g\in \mathbb{F}_2[x]$ of degree $n$ with $g(0) \neq 0$. Therefore, the order of its corresponding companion matrix satisfies $ord(C(g))=ord(g)=2^n -1$ and $\det(C(g)) = (-1)^n$. We use this companion matrix $C(g)$ that is a binary low dense matrix of maximum order $2^n -1$
to construct a nonlinear function $F$ for our proposed scheme. For simplicity we write $C(g)=\mathbf{U}$ in the sequel. Then, the set $\mathrm{S}=\{\mathbf{I}, \mathbf{U}, \mathbf{U}^2, \ldots, \mathbf{U}^{2^{n}-2}\}$ contains $2^n -1$ invertible binary matrices of dimension $n\times n$.

The mapping $F$ is obtained using  $2^d$ linear transformations from the set $\mathrm{S}$  indicated by the linear transformations $F_0, \ldots, ,F_{2^{d}-1}$.
As a consequence, $F:\mathbb{Z}^{n+d}\longrightarrow \mathbb{Z}^{n}$ is a mapping  defined by $F\big(\mathbf{a}, \mathbf{b}\big)=F_{\alpha}(\mathbf{a})=\mathbf{a}\mathbf{U}^{\alpha}$, where $\alpha=\sum_{i=0}^{d-1}{b_i 2^i}\in \{0, 1, \ldots, 2^{d}-1\}$ is determined by the vector $\mathbf{b}$ as the control line of the multiplexer. The public set $\mathrm{S}$ and  a secret control line $\mathbf{b}$ of the multiplexer define
this function.
On the other hand, the output $F(\mathbf{a}, \mathbf{b})=\{f_1(\mathbf{a}, \mathbf{b}), \ldots, f_n(\mathbf{a}, \mathbf{b})\}$ is  the concatenation of the output of $2^d$ linear operators selected according to the  logic of multiplexer \cite{Function-f}.
Each component function can be written mathematically as
\begin{align}\label{function-F}
f_i(\mathbf{a}, \mathbf{b})\equiv \mathop  \bigoplus \nolimits_{j  = 0}^{{2^d} - 1} {D_j }(\mathbf{b}){\mathbf{u}_{ji}}(\mathbf{a}) ~ (\textrm{mod} ~2),
\end{align}
where $\mathbf{u}_{ji}$ is the $i$th row of the matrix $\mathbf{U}^{j}$, and $D_j(\cdot)$ is a function defined as $D_j(\mathbf{b})=D_\sigma(\mathbf{b})= ({{\bar i}_1} \oplus {b_1}) \cdots ({{\bar i}_d} \oplus {b_d}) $ over $ \sigma  = ({i_1}\,, \ldots ,\,{i_d})$ which is the binary representation of $j$
and $\bar v$ is the complement of $v$.
In a similar way  \cite{Function-f}, it can be proved that the algebraic degree of each component function of the proposed mapping $F$ and their nonzero linear combinations is $d+1$.
\subsection{Key generation}\label{key generation}
Encryption is done using the following secret keys that are chosen by the authorized transmitter and receiver.
\begin{enumerate}
  \item A random regular $(n=n_0b, k=(n_0-1)b, d_c)$-QC-LDPC code constructed from RDFs  with a parity-check matrix of  size  $(n-k)\times n$ in the form of (\ref{parity check}) and constant row weight $d_c$.
    This RDF-QC-LDPC code with the parity-check matrix $\mathbf{H}_{qc}$ is used to construct a QC-LDPC lattice (according to the previous section).
\item A vector $\mathbf{s}$ as $l_1$-bit initial value (seed) of a linear feedback shift register (LFSR)  which is described using a polynomial $q$. We use this LFSR to produce an $n$-tuple intentional error vector $\mathbf{e}$ in the scheme.
  	For reducing the key size  and increasing the period of the resultant keystream,  we use a reseeding mechanism for this LFSR at the end of each period. This mechanism uses the modular division circuit with a different polynomial $p$  for another LFSR of the same length  proposed in \cite{Modulation-circuit}. With a proper selection of the polynomials $q$ and $p$, the period of the LFSR is equal to $(2^{l_1} - 1)^2$.
Since, the output of each period of the LFSR is a vector with approximately $(2^{l_1} - 1)$ bits, then the length of the LFSR should be selected in such a way that the message length $n$ is close to the period of LFSR, that is, $(2^{l_1} - 1)\approx n$.
Therefore, we consider $l_1=\lceil \log_2 n \rceil$ which implies $(2^{l_1} - 1)^2\approx n^2$. Using this procedure, we can produce $n$ different pseudorandom binary vectors $\mathbf{e}_i$, $i=1, \ldots, n$, of length  $n$ for the encryption algorithm.
\item An $n\times n$ block diagonal permutation matrix $\mathbf{P}=\textrm{diag}(\boldsymbol{\pi}_1 ,\ldots ,\boldsymbol{\pi}_v)$ formed by ${q\times q}$ sub-matrices  $\boldsymbol{\pi}_i$, for $i=1, \ldots, v$,
    where $v=n/q$.
    The diagonal elements $\boldsymbol{\pi}_i$'s are permutation sub-matrices, so the Hamming weight of each row and column is one.
    We control $q\times q$   sub-matrices  $\boldsymbol{\pi}_i$, for $i=1, \ldots, v$,  by $v$ different initial values for an LFSR of length $ \lceil\log_2 q\rceil$. We describe it in Section \ref{sec-key-size}. Since, the  sub-matrices $\boldsymbol{\pi}_i$s are saved instead of  the  permutation matrix $\mathbf{P}$, it is  needed to save only the corresponding initial value of the LFSR.
    We concatenate these initial values and store them in a vector $\mathbf{t}$ of length $l_4=v \lceil\log_2 q\rceil$ bits as a key.
 \item An $l_2$-bit initial vector $\mathbf{h}$  of an LFSR that controls the multiplexer in the construction of the nonlinear mapping $F$.  The vector $\mathbf{h}$ serves as the control line and the multiplexer outputs one of the outputs of the linear transformations according to this control line.
\end{enumerate}

Let $d_v$ denote the number of nonzero elements in each row/column of $\mathbf{H}_i$ in $\mathbf{H}_{qc}$, thus the row weight of $\mathbf{H}_{qc}$ is $d_c=n_0d_v$.
Let $\mathbf{H}_{n_0-1}$ be non-singular, particularly, this implies that $d_v$ is odd. Then, the systematic generator matrix of  this QC-LDPC code is
%
\begin{eqnarray}\label{G-qc}
 \mathbf{G_{\mathcal{C}}}=\left[\begin{array}{c|c}
            \mathbf{I_{k}}\qquad &  \begin{array}{c}
                                 (\mathbf{H}_{n_0-1}^{-1}\mathbf{H}_0)^T \\
                                 (\mathbf{H}_{n_0-1}^{-1}\mathbf{H}_1)^T \\
                                 \vdots\\
                                 (\mathbf{H}_{n_0-1}^{-1}\mathbf{H}_{n_0-2})^T
                                \end{array}
                                 \end{array}\right]_{k\times n},
\end{eqnarray}
where $[\, \cdot \,]^T$ denotes the transposition operation.
Then, the generator matrix $\mathbf{G_{\Lambda}}$ of the corresponding QC-LDPC lattice is obtained by replacing it in Eq. (\ref{generator matrix}).

Since each circulant matrix $\mathbf{H_i}$, for $i=0, \ldots, n_0-1$,  is completely described by its first row, the QC-LDPC code with parity-check matrix (\ref{parity check}) is given by the first row of these circulant blocks. Therefore,  we save the first row of $\mathbf{H}_{qc}$ instead of the entire matrix.


\subsection{Encryption Algorithm}
To encrypt  a message $\mathbf{m} \in \mathbb{Z}^{n}$, an intentional pseudorandom error vector  $\mathbf{e}\in \mathbb{F}_2^{n}$ is generated using  the LFSR with the secret initial value $\mathbf{s}$ and the reseeding mechanism.
This error vector of length $n$ has an arbitrary Hamming weight.
%
Then, the ciphertext is computed as follows
\begin{align*}
  \mathbf{y}&=\Big(2F\big((\mathbf{m}+\mathbf{\overline{e}}), \mathbf{h}\big) \mathbf{G_{\Lambda}}-\mathbf{1}+2\mathbf{e}\Big)\mathbf{P}\\
  &=\Big(2\mathbf{x} \mathbf{G_{\Lambda}}-\mathbf{1}+2\mathbf{e}\Big)\mathbf{P},
\end{align*}
where $\mathbf{\overline{e}}$ is the complement of the vector $\mathbf{e}$ in $\mathbb{F}_2$.
The nonlinear function $F$ maps the intentionally corrupted message $(\mathbf{m}+\mathbf{\overline{e}})$ to the vector $\mathbf{x}=F\big((\mathbf{m}+\mathbf{\overline{e}}), \mathbf{h}\big)=(\mathbf{m}+\mathbf{\overline{e}})\mathbf{U}^{\alpha}$, in which $\alpha=\sum_{i=0}^{d-1}{h_i 2^i}$ and  $\mathbf{h}$ is the secret control line of the employed  multiplexer in $F$.  Indeed, the ciphertext $\mathbf{y}$ is the permutation of  the lattice point $\boldsymbol{\tilde{\lambda}}=2\mathbf{x} \mathbf{G_{\Lambda}}-\mathbf{1}$ which has been perturbed by the vector $2\mathbf{e}$.

In general, the $j$th instance of the ciphertext $ \mathbf{y}_j$ ($j\geq 1$) corresponds to the $j$th instance of the plaintext $ \mathbf{m}_j$  as follows
\begin{align*}
  \mathbf{y}_j&=\Big(2F\big((\mathbf{m}_j+\mathbf{\overline{e}}_{(j)_{N_e}}), \mathbf{h}_{(j)_{N_h}}\big) \mathbf{G_{\Lambda}}-\mathbf{1}+2\mathbf{e}_{(j)_{N_e}}\Big)\mathbf{P}_{(j)_{N_p}},
\end{align*}
where $(j)_{N}$ is considered as $j \mod N$. The numbers $N_e, N_h$ and $N_p$ express the total possibilities of the intentional error vector $\mathbf{e}$, the multiplexer select logic $\mathbf{h}$ and the permutation matrix $\mathbf{P}$, respectively.
Indeed, the vectors $\mathbf{e}$ and $\mathbf{h}$ and the permutation matrix $\mathbf{P}$ are changed corresponding to the output of the used LFSRs in their producing process. For simplicity, in the following, we ignore their subscripts.

As each lattice point $\boldsymbol{\tilde{\lambda}}=2\mathbf{x} \mathbf{G_{\Lambda}}-\mathbf{1}$ is a vector with odd components,
we must add $2\mathbf{e}$ instead of $\mathbf{e}$ as intentional error vector in the encryption process. Otherwise,
 $2\mathbf{x} \mathbf{G_{\Lambda}}-\mathbf{1}+\mathbf{e}$ is a vector with some even components that reveals the perturbed positions.
 Moreover, based on Section \ref{LDPC lattice encoding}, each lattice point in $\Lambda(\mathcal{C})$ can be expressed  as a vector of the form
$\mathbf{c}+4\mathbf{z}$, where $\mathbf{c}\in \mathcal{C}$ (converted into $\pm 1$) and $\mathbf{z}\in \mathbb{Z}^n$. When we add the  vector $2\mathbf{e}$ to a lattice point in the encryption process, we get
\begin{equation*}
 \mathbf{c}+4\mathbf{z}+2\mathbf{e}= \mathbf{c'}+4\mathbf{z'},
\end{equation*}
where $\mathbf{c'}$ is a vector with components $\pm 1$ and $\mathbf{z'}\in \mathbb{Z}^n$. Hence, $\boldsymbol{\tilde{\lambda}}+2\mathbf{e}$ is another lattice point that gives no information about the intentional error vector $\mathbf{e}$.
%
\subsection{Decryption Algorithm}
For decryption, the authorized receiver must be aware of the intentional error vector $\mathbf{e}$  that the transmitter uses by encrypting each plaintext.
In addition, both the authorized receiver and transmitter need to use the same LFSR with the same seed $\mathbf{s}$ as a part of the secret key.
 Therefore, if they use the modular division circuit for generating the pseudorandom error vector $\mathbf{e}$ simultaneously, they can use the same error vector  $\mathbf{e}$ for encryption and decryption. Hence, the decryption is done by employing the  following steps:
 \begin{enumerate}
   \item Multiply the ciphertext $\mathbf{y}$ by $\mathbf{P}^{-1}=\mathbf{P}^{T}$ and get
   \begin{equation*}
   \mathbf{y'}= \mathbf{y}\mathbf{P}^{T}=2\mathbf{x} \mathbf{G_{\Lambda}}-\mathbf{1}+2\mathbf{e}.
   \end{equation*}
   \item
Subtract $2\mathbf{e}$ from $\mathbf{y'}$ to get $2\mathbf{x} \mathbf{G_{\Lambda}}-\mathbf{1}$.
   \item Recover the vector $\mathbf{x}=F\big((\mathbf{m}+\mathbf{\overline{e}}), \mathbf{h}\big)$  by adding the vector $\mathbf{1}$ to $\mathbf{y'}-2\mathbf{e}$ and multiplying the result by $\frac{1}{2}\mathbf{G^{-1}_{\Lambda}}$, where
    \begin{equation*}
    \mathbf{G^{-1}_{\Lambda}}=\left[\begin{array}{cc}
   \mathbf{I_{k}}& -\frac{1}{2}\mathbf{A_{k\times (n-k)}} \\
    \mathbf{0_{(n-k)\times k}} & \frac{1}{2}\mathbf{I_{(n-k)}}
    \end{array}\right].
    \end{equation*}
   \item Apply the inverse of the function $F$ on $\mathbf{x}$ using the secret vector $\mathbf{h}$ and recover the intentionally corrupted message vector $\mathbf{m}'=\mathbf{m}+\mathbf{\overline{e}}$.
    \item   Retrieve the original message $\mathbf{m}$ by computing $\mathbf{m}'-\mathbf{\overline{e}}$.
 \end{enumerate}

We can use this system for communication in which the ciphertext is transmitted over a noisy channel.
This entails some modifications to our cryptosystem to present a joint scheme that can process encryption, channel coding and modulation in a single step which is discussed in the next section.
\section{The proposed Joint Encryption, channel coding and modulation scheme}~\label{sec:joint encryption}
In this section, we use QC-LDPC lattices to introduce a joint  scheme to provide secure communication  over bandwidth-limited (high-SNR) AWGN channels.

For communications over a noisy power constrained AWGN channel, the encoding operation
must be accompanied by a shaping method.
This prevents the transmission power of a codeword from being unnecessarily increased. Indeed, we make sure that only lattice points that belong to a shaping region are actually used. Then, instead of mapping the message vector $\mathbf{m}$ to the lattice point $\mathbf{m}\mathbf{G_{\Lambda}}$ in $\Lambda$, it should be mapped to another lattice point $\mathbf{m'}\mathbf{G_{\Lambda}}$, belonging to the shaping region.

Some known shaping methods in the literature are hypercube shaping, Voronoi  shaping and spherical shaping. In theoretical approaches, the infinite lattice  is intersected with a spherical shaping to produce a power-constrained lattice code.
Due to the high computational complexity  of this shaping, we consider an efficient hypercube shaping algorithm  that has minimum complexity among other shaping methods for QC-LDPC lattices  to obtain finite lattice constellations \cite{Full-Duplex}.

Thus, we choose a signal constellation formed by a QC-LDPC lattice together with a hypercube shaping region for  process of the mapping. Indeed, each message is modulated to one of the constellation points  using QC-LDPC lattice encoding and a suitable shaping method.
\subsection{The Proposed Encryption Algorithm}~\label{encrption-joint}
The ciphertext of the proposed symmetric key encryption scheme
is in the following form:
	\begin{equation}\label{joint-encryption-eq}
	\mathbf{y}=\Big(2F\big((\mathbf{m}+\mathbf{\overline{e}}), \mathbf{h}\big) \mathbf{G_{\Lambda}}-\mathbf{1}+2\mathbf{e}\Big)\mathbf{P}.
	\end{equation}
All  operations, such as applying the function $F$ on $\mathbf{m}'=\mathbf{m}+\mathbf{\overline{e}}$ and  computing  the  lattice point $\boldsymbol{\tilde{\lambda}}=2\mathbf{x} \mathbf{G_{\Lambda}}-\mathbf{1}$, and so on, are computed over the real numbers $\mathbb{R}$. Therefore, the resulting ciphertext $\mathbf{y}$ may have large components,  even if we restrict the message components.
While transmitting the ciphertext over a noisy power constrained AWGN channel, the encrypting operation must be accompanied by a shaping method. In this way,
the vector $\boldsymbol{\tilde{\lambda}}$ is limited to a region around the origin that leads to the reduction of  ciphertext transmission power.

On the other hand, the matrices $\mathbf{U}^{i}$s ($i=0, \ldots, 2^n-2$) in the set $S$ have no structure and may be dense enough. We have observed that some columns in  $\mathbf{U}^{i}$s are all one vectors. Therefore,  even by applying a shaping method on $\mathbf{x}=F(\mathbf{m}', \mathbf{h})=\mathbf{m}'\mathbf{U}^{\alpha}$, where  $\alpha=\sum_{i=0}^{d-1}{h_i 2^i}$,  the vector $\mathbf{x}$ may be mapped into  a big hypercube with high complexity.
Thus, we restrict the components of the input integer vector $\mathbf{m}$ to the following finite constellation before shaping:
\begin{eqnarray}\label{constelation}
	m_{2i} \in \left\{x\in\mathbb{Z}\,\left|\, -L_{i}\leq x\leq -1 \right.\right\}, \,\,\, i=1,\ldots, n/2,\\
    m_{2i-1} \in \left\{x\in\mathbb{Z}\,\left|\, 0\leq x\leq L_{i}-1 \right.\right\}, \,\,\, i=1,\ldots, n/2,\nonumber
	\end{eqnarray}
where $L_i$'s are positive integers. In this way, $-L_{i}\leq m'_{2i}\leq 0$ and $0\leq m'_{2i-1}\leq L_{i}$, for $i=1,\ldots, \frac{n}{2}$.
Therefore,  the vector $\mathbf{x}$  in the worst case lies in an $n$-dimensional hypercube around the origin such that  $x_{i}\in \{x\in\mathbb{Z}\, | \, -\frac{n}{2}L_{i} \leq x\leq \frac{n}{2}L_{i} \}$.

In our proposed scheme, it is desirable to design an optimal finite constellation for the entries of $\mathbf{m}$ such that the  sizes of the entries of $\mathbf{x}=\mathbf{m}'\mathbf{U}^{\alpha}$ become as small as possible. Designing such constellation could be an interesting problem for future work.

In the next step,  we   compute  the  lattice point $\boldsymbol{\lambda}=\mathbf{x}\mathbf{G_{\Lambda}}$ in the encryption process. Since all operations are performed over  real numbers,  the entries of  $\mathbf{x}\mathbf{G_{\Lambda}}$  can  also be large.
 Therefore, to make the scheme practical, we use a hypercube shaping method to  keep the corresponding entries of the ciphertext vector as small as possible.
Indeed,  instead of mapping the vector $\mathbf{x}$ to the lattice point  $\boldsymbol{\lambda}=\mathbf{x}\mathbf{G_{\Lambda}}$ in the infinite lattice $\Lambda$, it is mapped to a lattice point $\boldsymbol{\lambda}'=\mathbf{x'}\mathbf{G_{\Lambda}}$
inside an $n$-dimensional  hypercube such that  $|\lambda_i'|\leq nL_i-1$, for $i=1,\ldots, n$.

In the first step we enforce that $-\frac{n}{2}L_{i}\leq x_i\leq \frac{n}{2}L_{i}$, for $i=1,\ldots ,n$. Therefore,
the needed restriction condition of the input vector to a finite constellation  for the shaping algorithm is satisfied.
Thus, we consider a new lattice point
\begin{equation}\label{shaped-vector}
\boldsymbol{\lambda}'=\mathbf{x'}\mathbf{G_{\Lambda}}=(\mathbf{x}-\mathbf{z}\mathbf{L})\mathbf{G_{\Lambda}},
\end{equation}
instead of $\boldsymbol{\lambda}=\mathbf{x}\mathbf{G_{\Lambda}}$,
where $\mathbf{L}=\textrm{diag}(nL_1-1,\ldots ,nL_n-1)$ is an $n\times n$ diagonal matrix.
The  vector $\mathbf{z}$ is an integer vector of length $n$ that is chosen  such that the new lattice point components lie in an
$n$-dimensional hypercube around the origin \cite{Full-Duplex}.
To find the vector $\mathbf{z}$, we first solve the system of linear equations  obtained from (\ref{shaped-vector})  and then choose an integer $\mathbf{z}$   such that $|\lambda_i'|\leq nL_i-1$, for $i=1,\ldots ,n$. Therefore, we have
\begin{equation*}
\lambda_i'= \left\{ \begin{array}{cc}
                x_i-z_i(nL_i -1) & i=1,\ldots ,k, \\
                2(x_i-z_i(nL_i -1)) +\sum_{j=1}^{k}x'_j a_{j,i-k}, & i=k+1,\ldots ,n,
              \end{array}
              \right.
\end{equation*}
where  $a_{ji}$ is the $(j,i)$-th entry of the matrix $\mathbf{A}$ in $\mathbf{G_{\Lambda}}$.
We consider  $z_i=0$, for $1\leq i \leq k$ that leads to  $x'_i=x_i$ and $|\lambda_i'|=|x_i|\leq nL_i-1$.
Moreover, for $k+1\leq i \leq n$, we have
$-nL_i+1\leq 2(x_i-z_i(nL_i -1)) +\sum\nolimits_{j=1}^{k}x_j a_{j,i-k} \leq nL_i-1$,
or
$-\frac{1}{2}+\frac{x_i}{nL_i -1}+\frac{\sum_{j=1}^{k}x_j a_{j,i-k}}{2(nL_i -1)} \leq z_i \leq \frac{1}{2} +\frac{x_i}{nL_i -1} +\frac{\sum_{j=1}^{k}x_j a_{j,i-k}}{2(nL_i -1)}$.

The above interval contains only one integer number, thus it has the unique solution $z_i=\left\lfloor \frac{1}{nL_i-1} \left(x_i+ \frac{1}{2}\sum_{j=1}^{k}a_{j,i-k}x_j\right)\right\rceil$.

Indeed, we convert the vector $\mathbf{x}$ to $\mathbf{x'}=\mathbf{x}-\mathbf{z}\mathbf{L}$,
in order to embed the vector  $\boldsymbol{\lambda}'=\mathbf{x'}\mathbf{G_{\Lambda}}$  to  the following hypercube
\begin{eqnarray}\label{shaping-region}
\small
	\mathcal{L}= \left\{x\in\mathbb{Z}^n\left|\begin{array}{cc}
	-\frac{n}{2}L_{i}\leq x_i\leq \frac{n}{2}L_{i} &   i=1,\ldots, k, \\
	-nL_{i}+1 \leq x_i\leq nL_{i}-1 &   i=k+1,\ldots, n
	\end{array}
	\right.\right\}.
	\end{eqnarray}
Replacing $\mathbf{x}$ by the converted vector $\mathbf{x}'$, in the relation (\ref{joint-encryption-eq}), the transmitted vector over the noisy AWGN channel is expressed as $\mathbf{y}=(2 \mathbf{x'}\mathbf{G_{\Lambda}}-\mathbf{1}+2\mathbf{e})\mathbf{P}$.

\begin{figure}[h]
\centerline{\includegraphics[width=9in]{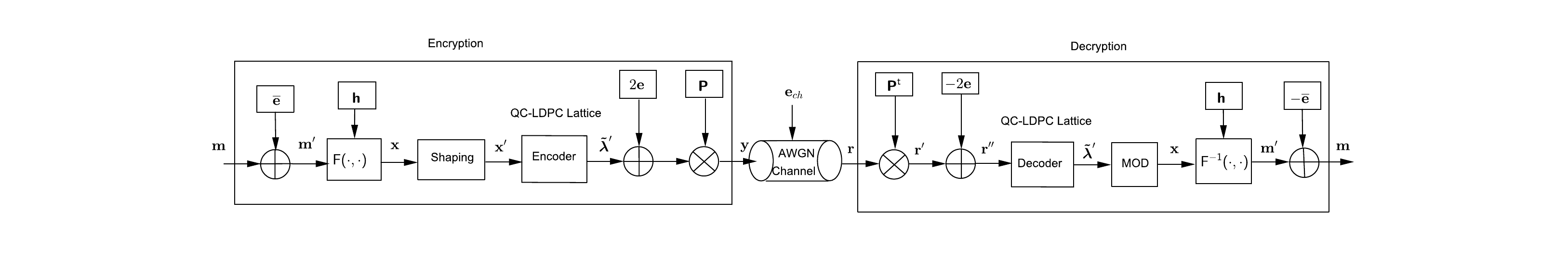}}
\caption{The proposed joint encryption, encoding and modulation scheme based on  QC-LDPC lattice codes}\label{block-diagram}
\end{figure}

The encryption algorithm is shown in  \figurename~\ref{block-diagram}.
The information rate of a lattice code $\Gamma$ of length $n$ (in bits/symbol) is defined to be $R=\log_2(|\Gamma|)/n$ \cite{LDLC}.
Since the points of the lattice code $\Gamma=\left(2\Lambda\cap\mathcal{L}\right) \mathbf{-1}$
are in bijective correspondence with the information integer vectors $\mathbf{m}$,  according to Eq.(\ref{constelation}), the information rate of this lattice code is  $R=\sum_{i=1}^{n}\log_2 (2L_i)/n$.

\subsection{The Proposed Decryption Algorithm}
The authorized receiver (Bob) attempts to decrypt the possibly erroneous received vector
$\mathbf{r}=(2 \mathbf{x'}\mathbf{G_{\Lambda}}-\mathbf{1}+2\mathbf{e})\mathbf{P}+\mathbf{e}_{ch}$,
where $\mathbf{e}_{ch}$ is the AWGN channel noise  that is drawn  from an i.i.d. Gaussian distribution with variance $\sigma^2$.

Using the secret keys $ \{\mathbf{s}, \mathbf{h}, \mathbf{H}_{qc}, \mathbf{t}\}$, the ciphertext is decrypted as follows:
\begin{enumerate}
   \item The received vector  $\mathbf{r}$ is multiplied by $\mathbf{P}^{-1}=\mathbf{P}^{T}$ to get
   \begin{equation*}
   \mathbf{r'}= \mathbf{r}\mathbf{P}^{T}=
   (2 \mathbf{x'}\mathbf{G_{\Lambda}}-\mathbf{1}+2\mathbf{e})+\mathbf{e}_{ch}\mathbf{P}^{T}.
   \end{equation*}
   \item Having the secret vector $\mathbf{s}$, the corresponding error vector  $\mathbf{e}$ and   $\mathbf{r''}=\mathbf{r'}-2\mathbf{e}$ are computed.
   \item  The vector $\mathbf{r''}=2\mathbf{x'} \mathbf{G_{\Lambda}}-\mathbf{1}+\mathbf{e}_{ch}\mathbf{P}^{T}$ is decoded using $\mathbf{H}_{qc}$ by applying the SPA iterative decoding algorithm of QC-LDPC lattices \cite{QC-LDPC-Lattices}. Then  $\boldsymbol{\tilde{\lambda}}'=2\mathbf{x'}\mathbf{G_{\Lambda}}-\mathbf{1}$ is obtained.
    \item The vector $\mathbf{x}$ is recovered from the shaped lattice point  $\boldsymbol{\tilde{\lambda}}'$ using Algorithm \ref{unshape}.

\begin{algorithm}
 \begin{algorithmic}[1]
 \small
 \Procedure{MOD}{$\boldsymbol{\tilde{\lambda}}', (L_1,\ldots ,L_n), \mathbf{G^{-1}_{\Lambda}}$}
 \State $\mathbf{x'} \gets \left\lfloor\left(\frac{\boldsymbol{\tilde{\lambda}}'+\mathbf{1}}{2}\right)\mathbf{G^{-1}_{\Lambda}}\right\rceil$
 \For{$i=1:n$}
  \If{$\bmod(x'_i, (nL_i-1))< \frac{n}{2}L_{i}$}
   \State $r_i\gets \bmod(x'_i, (nL_i-1))$
   \Else{}
  \State $r_i\gets \bmod(x'_i, (nL_i-1))-(nL_i-1)$
  \EndIf
 \EndFor
 \State \textbf{return} $\mathbf{x}=(r_1,\ldots ,r_n)$.
 \EndProcedure
 \end{algorithmic}
 \caption{Recover original vector}
 \label{unshape}
\end{algorithm}
\normalsize

\item Retrieving the vector $\mathbf{m'}=\mathbf{m}+\mathbf{\overline{e}}$ from the vector   $\mathbf{x}=\mathbf{m'}\mathbf{U}^{\alpha}$ is equivalent to computing $\mathbf{m'}=\mathbf{x}(\mathbf{U}^{\alpha})^{-1}$ by the secret vector $\mathbf{h}$,  in which $\alpha=\sum_{i=0}^{d-1}{h_i 2^i}$. Then, subtracting the vector $\mathbf{\overline{e}}$ from  $\mathbf{m'}$ recovers  the original message $\mathbf{m}$.
\end{enumerate}

\section{Efficiency}~\label{sec:Key Size and Complexity}
The efficiency of the cryptosystem is measured in terms of the key size and the computational complexity of encryption and decryption processes.
\subsection{Complexity}\label{complexity}
Encryption is performed by computing $\mathbf{m}'=\mathbf{m}+\mathbf{\overline{e}}$, then mapping it to the vector $\mathbf{x}=\mathbf{m}'\mathbf{U}^{\alpha}$ by the function $F$  through a multiplexer controlled by the binary vector $ \mathbf{h}$.
The vector $\mathbf{x}$ is converted to the vector $\mathbf{x}'$  using a hypercube shaping method and then encoded by a QC-LDPC lattice to a lattice point $\boldsymbol{\tilde{\lambda}}'=2\mathbf{x'}\mathbf{G_{\Lambda}}-\mathbf{1}$.
By adding the intentional error vector $2\mathbf{e}$ and multiplying the whole combination by the permutation matrix $\mathbf{P}$, the overall system becomes non-systematic.
Therefore, an estimation of the computational complexity, caused by the encryption algorithm is given as
\begin{align*}\label{encryption complexity}
   C_{encrypt}=&   C_{add}(\mathbf{\overline{e}})+ C_{compute}(F(\mathbf{m}', \mathbf{h}))+C_{shaping}(\mathbf{x}\mathbf{G_{\Lambda}}) 
 + C_{encode}(\mathbf{x}')+C_{add}(2\mathbf{e})+ C_{product}\big(\mathbf{P}\big).
\end{align*}

%
In the encryption process, the terms $ C_{add}(\mathbf{\overline{e}})$, $C_{add}(2\mathbf{e})$ and $C_{product}\big(\mathbf{P}\big)$ have  lower order complexity, which are linear, than other terms.
Indeed,  the complexity of the encryption process is upper-bounded by the complexity of
$ C_{compute}(F((\mathbf{m}+\mathbf{\overline{e}}) , \mathbf{h}))+C_{shaping}(\mathbf{x}\mathbf{G_{\Lambda}})+C_{encode}(\mathbf{x}')$.

The time and space complexities of QC-LDPC lattice encoding are linear in terms of the dimension of lattice \cite{QC-LDPC-Lattices}.
The overall computational complexity of the hypercube shaping for a $\boldsymbol{\lambda}=\mathbf{x}\mathbf{G_{\Lambda}} \in \Lambda$ is $O(n w_c)$, where $w_c$ is the average number of nonzero elements in a row of $\mathbf{G_{\Lambda}}$ \cite{Hassan}.
Since $F(\mathbf{m}',  \mathbf{h})=\mathbf{m}'\mathbf{U}^{\alpha}$, where $\alpha=\sum_{i=0}^{d-1}{h_i 2^i}$, then $C_{compute}(F(\mathbf{m}',  \mathbf{h}))$  is equal to the computational complexity of vector $\mathbf{m}'=(\mathbf{m}+\mathbf{\overline{e}})$ times matrix $\mathbf{U}^{\alpha}$.
The function can be implemented using a similar pipelined architecture proposed in \cite{Function-f}. Let $\alpha=h_0+2h_1+2^2 h_2+\cdots+2^{d-1}h_{d-1}$, then   $\mathbf{U}^{\alpha}=\mathbf{U}^{h_0}\cdot(\mathbf{U}^2)^{h_1}\cdot(\mathbf{U}^{2^2})^{h_2} \cdots (\mathbf{U}^{2^{d-1}})^{h_{d-1}}$, where $h_i$s are $0$ or $1$.
Indeed, the proposed function can be implemented using the matrices $\mathbf{U}, \mathbf{U}^2, \ldots, \mathbf{U}^{2^{d-1}}$ and $d$ number of 2-to-1 multiplexers instead of all matrices ($2^d$ numbers) in $S$ and one $2^d$-to-$1$ multiplexer.
 Therefore, its computational complexity is of order  $O(n^2)$.
As a consequence,  the complexity of the encryption process is of order $O(n^2)$, in terms of lattice dimension $n$.
Since, we have $d$ stages in the pipeline, the latency of the architecture is $O(d)$.
Using similar arguments, the decryption complexity is expressed as
\begin{align*}
C_{decrypt}=&C_{product}\big(\mathbf{P}^T\big)+C_{add}(-2\mathbf{e})+C_{decode}(\mathbf{r''}) 
+ C_{MOD}(\boldsymbol{\tilde{\lambda}}')+C_{compute}(F^{-1}(\mathbf{x}))+C_{add}(-\overline{\mathbf{e}}).
\end{align*}
Since the dominant terms that have much larger effect on the implementation complexity are $C_{decode}(\mathbf{r''})$, $ C_{MOD}(\boldsymbol{\tilde{\lambda}}')$ and $C_{compute}(F^{-1}(\mathbf{x}))$, the decryption complexity is upper-bounded by the complexity of  $C_{decode}(\mathbf{r''})+ C_{MOD}(\boldsymbol{\tilde{\lambda}}')+C_{compute}(F^{-1}(\mathbf{x}))$.

The QC-LDPC lattice decoding  has computational complexity of $O(n d_v  I)$, where $I$ is the maximum number of iterations required by the decoding algorithm to correct the error and $d_v $ is the column weight of the parity-check matrix $\mathbf{H}_{qc}$.
Hence, QC-LDPC lattices have linear computational complexity in the lattice dimension \cite{QC-LDPC-Lattices}.

The information integer components $x_i$ are recovered from $x'_i$ after multiplication by the matrix $\mathbf{G^{-1}_{\Lambda}}$ followed by a simple modulo operation that is explained in Algorithm \ref{unshape}. 
Therefore, $C_{MOD}(\boldsymbol{\tilde{\lambda}}')=C_{mult}(\mathbf{G^{-1}_{\Lambda}})$, where according to the structure of $\mathbf{G^{-1}_{\Lambda}}$, it can be demonstrated that it is equal to complexity of the encoding algorithm of QC-LDPC lattices which is  linear in terms of lattice dimension.
The computational complexity of applying  $F^{-1}(\mathbf{x})$ is equivalent to multiplication of the matrix $(\mathbf{U}^{\alpha})^{-1}$ to the vector $\mathbf{x}$. In the same way, its computational complexity is of order  $O(n^2)$. Hence, the  total computational  complexity  for decryption is bounded by $O(n^2)$.

The complexity of computing the nonlinear  function $F$ and  $F^{-1}$ is a bottleneck for our scheme. Indeed, we had to sacrifice the linear complexity of our design to prevent some chosen plaintext attacks like differential attack against the proposed scheme. However, designing another nonlinear function which can be implemented with linear complexity is desirable and open for further research.

\subsection{Message expansion}
According to Section \ref{encrption-joint}, the shaped vectors $2\mathbf{x'} \mathbf{G_{\Lambda}}-\mathbf{1} \in 2\Lambda-\mathbf{1}$ are uniformly distributed over the hypercube $2\mathcal{L}-\mathbf{1}$, where $\mathcal{L}$ is presented in Eq. (\ref{shaping-region}).
Therefore, the ciphertext $\mathbf{y}=(2 \mathbf{x'}\mathbf{G_{\Lambda}}-\mathbf{1}+2\mathbf{e})\mathbf{P}$ belongs to the following set
\small
\begin{eqnarray}\label{Hypercube-region}
	 \left\{x\in\mathbb{Z}^n\left|\begin{array}{cc}
	-nL_{i}-1\leq x_i\leq nL_{i}+1 &   i=1,\ldots, k, \\
	-2nL_{i}+1 \leq x_i\leq 2nL_{i}-1 &   i=k+1,\ldots, n
	\end{array}
	\right.\right\}.
	\end{eqnarray}
\normalsize
For simplicity, we consider $L_i = L$, for $i=1,\ldots, n$. Then, the information rate of our cryptosystem is $R=\log_2(2L)$.
Furthermore, according to  Eq. (\ref{constelation}), the integer vector $\mathbf{m}$ is restricted to the finite constellation $m_i \in \left\{-L,\ldots, L-1 \right\}$, for $i=1,\ldots, n$. Therefore, the plaintext size is at most $n\lceil\log_2(2L)\rceil$ bits and  the number of bits required to derive a ciphertext is $(n-k)\lceil\log_2(4nL-1)\rceil +k\lceil\log_2(2nL+3)\rceil$.
Therefore, the message expansion of our cryptosystem is
\begin{IEEEeqnarray*}{rCl}\label{message expansion}
\small
\frac{(n-k)\lceil\log_2(4nL-1)\rceil +k\lceil\log_2(2nL+3)\rceil}{n\lceil\log_2(2L)\rceil}.
\end{IEEEeqnarray*}
If we consider the parameters $n_0=6$, $d_v=3$ and $b=43$ that introduce a QC-LDPC lattice with  $(k, n)=(215, 258)$,
this ratio approaches $1$ as $L$ approaches  infinity. Moreover, the message expansion of our cryptosystem, belongs to the interval $[1, 5.6]$, for $L\geq 2$ and the proposed parameters.


\subsection{Key size}\label{sec-key-size}
The secret key that needs
to be exchanged between the sender and authorized receiver consists of an initial value $\mathbf{s}$ of the $l_1$-bit LFSR for generating the error vector  $\mathbf{e}$,  the  parity-check matrix $\mathbf{H}_{qc}$ for  decoding of QC-LDPC lattices,  the $d$-bit vector $\mathbf{h}$ that determines the selection logic of the multiplexer for nonlinear mapping $F$, and the vector $\mathbf{t}$ correspond to the permutation matrix $\mathbf{P}$.

In the proposed encryption scheme, the random error vector $\mathbf{e}$ is  generated using an $l_1$-bit LFSR along with a modular division circuit proposed  in \cite{Modulation-circuit}. Therefore, we store the initial vector  $\mathbf{s}$  of the LFSR that requires a memory of approximately $l_1=\lceil\log_2 n\rceil$ bits.

The required memory for saving the vector $\mathbf{h}$
is $l_2$ bits. We consider it in order for the cryptosystem to be secure against different cryptanalysis.
Here we consider it of size approximately $7\lceil\log_2 n\rceil$.

To save the secret key $\mathbf{H}_{qc}$, we can just save the nonzero positions of the first rows of $(\mathbf{H}_i)_{b\times b}$, for $i=0, \ldots, n_0-1$, and keep it as the secret parity-check matrix.
Thus, its storage involves at most $l_3=d_v\lceil\log_2 b\rceil n_0$ memory bits, where $d_v$ is the row/column  Hamming weight of $\mathbf{H}_i$, for $i=0, \ldots, n_0-1$.

We use an efficient hardware structure for the permutation of a vector that is based on a shift register and some multiplexers  \cite{Deepthi-Nonlinear}.
By this method,  for different message blocks, we generate different permutation matrices.  Under this low hardware complexity, we can reduce the key size of the scheme compared to other joint schemes as well as be secure against known attacks.
\begin{figure}[h]
	\centerline{\includegraphics[width=4in]{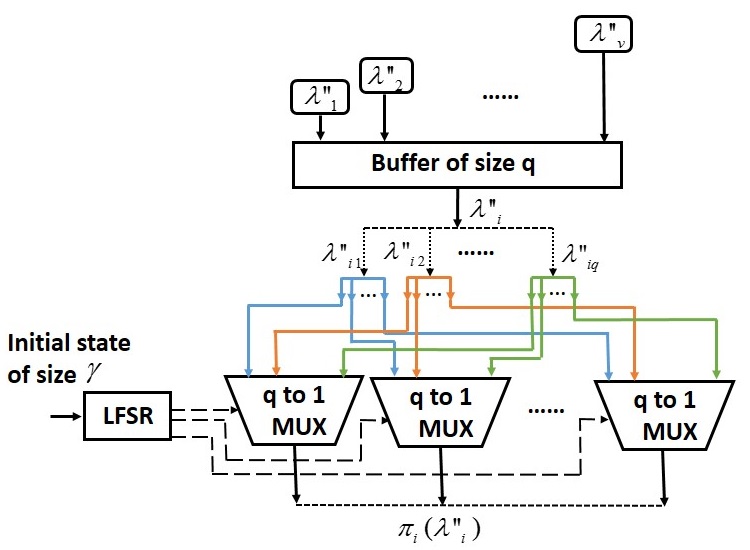}}
	\caption{Hardware design for vector permutation by the matrix  $\mathbf{P}$.}\label{Permutation-implement}
\end{figure}
The designed hardware structure for permuting a vector by $\mathbf{P}$, is shown in \figurename~\ref{Permutation-implement} that serves a buffer, an LFSR and some $q$-to-$1$ multiplexers.
In this structure, an LFSR of size $\gamma=\lceil\log_2 q\rceil$ generates data control line of the $q$-to-$1$ multiplexers.
For permuting the $n$-tuple vector $\boldsymbol{\lambda}''=2 \mathbf{x'}\mathbf{G_{\Lambda}}-\mathbf{1}+2\mathbf{e}$ in the last step of the encryption process, we divide it into $q$-tuple vectors $\boldsymbol{\lambda}''_i$, for $i=1, \ldots, v$.
Then, the first $q$-tuple vector $\boldsymbol{\lambda}''_1$ is stored in the buffer and is given as the input of the $q$-to-$1$  multiplexers.
Each component  of $\boldsymbol{\lambda}''_i$  belongs to the hypercube defined in Eq. (\ref{Hypercube-region}). Therefore, each component can be considered with maximum $r=\lceil\log_2(4nL-1)\rceil$   bits.
Then, we need $r$ multiplexers with $q$ inputs in order to the $r$ bits of each component of $\boldsymbol{\lambda}''_i$ be permuted simultaneously and result in the same component after permutation.
In this way, according to the control line generated by the LFSR, the multiplexers select one of the bits from the input vector $\boldsymbol{\lambda}''_{1_j}$, for $j=1, \ldots, q$, as the output. By the next output of the LFSR, the multiplexer selects another component of the input vector. Finally,  the $q$-tuple vector $\boldsymbol{\lambda}''_1$ is permuted at the end of
the first period of the LFSR which is  $2^{\gamma}-1= q$, where  $\gamma$ is the size of the initial vector. Then, the second $q$-tuple vector $\boldsymbol{\lambda}''_2$  is stored in the buffer and the initial value of the LFSR is changed. In the same way, $\boldsymbol{\lambda}''_2$  is permuted after reaching the period of the LFSR.
Therefore, each of  $\boldsymbol{\lambda}''_i$, for $i=1, \ldots, v$, is permuted at the end of a period of the LFSR.
Indeed, each permutation matrix $\boldsymbol{\pi}_i$, for $i=1, \ldots, v$, is controlled by the corresponding initial vector of the LFSR to determine the selection logic of the multiplexers. Since the $q\times q$   permutation  sub-matrices $\boldsymbol{\pi}_i$, for $i=1, \ldots, v$,  are saved instead of  the  permutation matrix $\mathbf{P}$, it is  needed to save only the corresponding initial vector of the LFSR.
Thus, the required memory bits for storing the  permutation matrix  $\mathbf{P}$  is $l_4=v \lceil\log_2 q\rceil$ bits.

Hence, the actual key length of the proposed cryptosystem is equal to $l_1+l_2+l_3+l_4=\lceil\log_2 n\rceil + 7\lceil\log_2 n\rceil +d_v\lceil\log_2 b\rceil n_0+ v \lceil\log_2 q\rceil$ bits.

For the proposed QC-LDPC lattice with  $(k, n)=(215, 258)$  and parameters $n_0=6$, $d_v=3$ and $b=43$, we can choose $q=43$ for the permutation matrix  $\mathbf{P}$ and  $l_2=61$ for the vector $\mathbf{h}$. Therefore, the key size of the proposed scheme is equal to $214$ bits. This key size is  small comparing with those of the proposed code and lattice based cryptography.

We summarize the  operation characteristics of the proposed
scheme in terms of  its parameters in \tablename~\ref{table2}.
\begin{table}[!ht]
\caption{Operation characteristics of the proposed
scheme.}
\centering
\begin{tabular}{|c||c|}
  \hline
Plaintext size (bit) & $n\lceil\log_2(2L)\rceil$  \\
  \hline
 Ciphertext size (bit) & $(n-k)\lceil\log_2(4nL-1)\rceil +k\lceil\log_2(2nL+3)\rceil$ \\
  \hline
\hspace{-0.4 cm} Key size (bit) &  $8\lceil\log_2 n\rceil +d_v\lceil\log_2 b\rceil n_0+ v \lceil\log_2 q\rceil$  \\
 \hline
Information rate & $R=\log_2(2L)$ \\
 \hline
Decryption Ops. & $O(n^2)$  \\
  \hline
Encryption Ops. &  $O(n^2)$ \\
  \hline
 \end{tabular}
 \label{table2}
 \end{table}
\section{Security of the proposed Scheme}~\label{sec:attacks}
In general, security of the RN-like cryptosystems is considered against four potential  attacks  reported  in the literature including brute force attack, differential-style attack \cite{Differential-Cryptanalysis}, Rao-Nam attack \cite{Rao-Nam}, Struik-Tilburg attack \cite{STRUIK} and message-resend attack \cite{Berson}.

The last three attacks are performed  based on the linearity of the encoding step in the encryption process.
They are not  applicable here because the use of the nonlinear function $F$ in the encryption algorithm prevents such attacks from being successful.

Rao-Nam attack is applied on any RN-like cryptosystems to estimate the encryption matrix from a large set of plaintext-ciphertext pairs \cite{Rao-Nam}. The security of the RN-like schemes against this kind of attacks depends mostly on the Hamming weight of the intentional error vectors.
This attack succeeds  only if the ratio of the Hamming weight of the intentional error vector over $n$ is small and it does not if its average Hamming weight is approximately  $  n/2$ \cite{Rao-Nam}.
According to the applied method for producing  the intentional error  $\mathbf{e}$ in our scheme, it has the Hamming weight  $  n/2$  on average.
In this method, the length of  $\mathbf{e}$ is equal to the period of the LFSR. Indeed, the output string (one cycle) of the LFSR  is considered as the intentional error  $\mathbf{e}$.
On the other hand, applying the nonlinear function $F$ on the plaintext before its encoding, does not allow  to estimate the encryption matrix from a large set of plaintext-ciphertext pairs.
Therefore, Rao-Nam attack can not be applied on the proposed scheme.

Struik-Tilburg attack requires enciphering an arbitrary message $\mathbf{m}$ until all distinct ciphertexts are obtained \cite{STRUIK}.
This attack is based on deriving the rows of the encryption matrix $\mathbf{G_{\Lambda}}$
by constructing unit vectors from the chosen plaintext or by solving a set of linear equations.
The proposed scheme is not vulnerable  to Struik-Tilburg attack and message-resent attack  \cite{Berson}, because the plaintext is transformed by means of an invertible and nonlinear function  $F$ before its encoding.

It is known that the security of nonlinear cryptosystems is determined by  the intentionally random error vectors and the nonlinear function $F$ \cite{Differential-Cryptanalysis}. Indeed, the encoding step and the permutation step increase further the security of the nonlinear system.
The main attack against nonlinear cryptosystems is  a differential-style attack that is a chosen plaintext attack \cite{Differential-Cryptanalysis}.  In the sequel, the security of the proposed cryptosystem is analyzed against Brute-Force and Differential attacks.

\subsection{Brute-Force Attacks}
The purpose of this attack is the enumeration of all possible secret keys $ \{\mathbf{s}, \mathbf{h}, \mathbf{H}_{qc}, \mathbf{t}\}$ in the proposed scheme until a meaningful message is obtained.
\begin{enumerate}
 \item The matrix $\mathbf{H}_{qc}$ is  the parity-check matrix (free of length-$4$ cycles) of an RDF-QC-LDPC code with code rate $R=(n_0- 1)/n_0$, code length $n=bn_0$ and column weight $d_v$. An attacker looks for the parity-check matrix corresponding to the used RDF-QC-LDPC code having public parameters $b, d_v, n_0$. For large enough parameters $b$ and $n_0$, it has been demonstrated  that there are a large number of different  RDF-QC-LDPC codes with the same code length, dimension and  row/column weight of the parity-check matrices \cite{Baldi-book}.
        The number of different QC-LDPC codes free of  length-$4$ cycles with the parameters $b, d_v, n_0$ that
        can be designed through the RDF-based approach, is lower bounded by  \cite[Theorem 4.12]{Baldi-book}
      \begin{align*}
      N_{RDF}(& b, d_v, n_0) \ge \frac{1}{b}{b\choose d_v}^{n_0}\prod\limits_{l = 0}^{{n_0} - 1} {\prod\limits_{j = 1}^{{d_v} - 1} {\frac{b}{{b - j}}} } 
     - \frac{{j\left[ {2 - b\,\bmod \,2 + ({j^2} - 1)/2 + l \cdot {d_v} \cdot ({d_v} - 1)} \right]}}{{b - j}}.
  \end{align*}
       For the proposed parameter $b=43, d_v=3$ and $n_0=6$, there are $2^{61}$ different  RDF-QC-LDPC codes with $n=258$ and code rate $R=5/6$.
 \item  We use an LFSR with $l_1=\lceil\log_2 n\rceil$-bit initial vector and a modular division circuit to generate the vector $\mathbf{e}$. With a suitable choice of the feedback polynomial used in the circuit,	the total choice of random	 vector $\mathbf{e}$ is   $(2^{l_1}-1)^2\approx (2^{\lceil\log_2 n\rceil})^2$. For our example, this number is approximately equal to  $2^{18}$.
  \item  The $l_2$-bit vector $\mathbf{h}$ determines the image of the nonlinear mapping $F$. The total  number of  the different vectors $\mathbf{h}$ is  $2^{l_2}$ that is $2^{61}$, for our cryptosystem.
    \item  Each permutation matrix $\boldsymbol{\pi}_i$, for $i=1, \ldots, v$, of the  block diagonal permutation matrices $\mathbf{P}$  is controlled by different  initial vectors of the LFSR  of size $\lceil\log_2 q\rceil$.
        The number of  different initial vectors for this LFSR is $2^{\lceil\log_2 q\rceil}$. Therefore, there are $2^{\lceil\log_2 q\rceil}$ different candidates for each $\boldsymbol{\pi}_i$, for $i=1, \ldots, v$. Hence, the total number of different  permutation matrices $\mathbf{P}$ is equal to $\left(2^{\lceil\log_2 q\rceil}\right)^v$, that is approximately $2^{36}$ for $q=43$ and $v=6$.
\end{enumerate}

Consequently, the complexity of the brute force attack is approximately $2^{176}$ which indicates a high order of security.
\subsection{Differential Cryptanalysis}
The $j$th  instance of the ciphertext in our cryptosystem is considered as
\begin{align*}
  \mathbf{y}_j&=\Big(2F\big((\mathbf{m}_j+\mathbf{\overline{e}}_{(j)_{N_e}}), \mathbf{h}_{(j)_{N_h}}\big) \mathbf{G_{\Lambda}}-\mathbf{1}+2\mathbf{e}_{(j)_{N_e}}\Big)\mathbf{P}_{(j)_{N_p}},
\end{align*}
which varies with the number of clock cycles.
The first round of the encryption is only used in the chosen plaintext attack proposed in \cite {Differential-Cryptanalysis}.
Indeed, the ciphertexts are computed for each message after resetting the encryption machine. Since, an attacker call $\mathbf{y}_1$ for different chosen plaintexts, the subscripts of $ \mathbf{m}_i$ and $\mathbf{y}_i$ are ignored for the time being and  $ \mathbf{m}^{(i)}$ and $ \mathbf{y}^{(i)}$ represent the plaintext and the corresponding ciphertext used in the $i$th call of $\mathbf{y}_1$. In our scheme, $ \mathbf{y}^{(\beta)}$, for $\beta=0, \ldots, 2^l$, is considered as the $\beta$th instance of the ciphertext corresponding to the plaintext $ \mathbf{m}^{(\beta)}$ as follows
\begin{equation*}
\mathbf{y}^{(\beta)}=
\Big(2\left(F\big((\mathbf{m}^{(\beta)}+\mathbf{\overline{e}}_1), \mathbf{h}_1\big) \right)'\mathbf{G_{\Lambda}}-\mathbf{1}+2\mathbf{e}_1\Big)\mathbf{P}_1.
\end{equation*}

This attack has three steps: estimate the permuted generator matrix $\mathbf{G_{\Lambda}}\mathbf{P}_1$, recover the intentional vector $\mathbf{e}_1$, and decode the ciphertext.
In the first step, the attacker computes the ciphertext corresponding to the first round of the encryption for the binary plaintext $ \mathbf{m}^{(\beta)}$, for $\beta=0, \ldots, 2^l$, in which the last $(n-k)$ bit of message is equal to zero.
Since,
\begin{align*}
\mathbf{y}^{(\beta)}&=\Big(2 \mathbf{x'}^{(\beta)}\mathbf{G_{\Lambda}}-\mathbf{1}+2\mathbf{e}_1\Big)\mathbf{P}_1\\
&=\Big(2 (\mathbf{x}^{(\beta)}-\mathbf{z}\mathbf{L})\mathbf{G_{\Lambda}}-\mathbf{1}+2\mathbf{e}_1\Big)\mathbf{P}_1,
\end{align*}
the attacker computes
\begin{align*}
\mathbf{y}^{(\beta)} ~ (\textrm{mod}~ \mathbf{L})&=\left(2 \mathbf{x}^{(\beta)}\mathbf{G_{\Lambda}} -\mathbf{1}+2\mathbf{e}_1\right)\mathbf{P}_1\\
&=\left(2F((\mathbf{m}^{(\beta)}+\mathbf{\overline{e}}_1), \mathbf{h}_1) \mathbf{G_{\Lambda}}-\mathbf{1}+2\mathbf{e}_1\right)\mathbf{P}_1.
\end{align*}
Then, for $\beta=0, \ldots, 2^l$, the attacker has
\begin{align*}
\mathbf{v}^{(\beta)}=\dfrac{\mathbf{y}^{(\beta)}~ (\textrm{mod}~ \mathbf{L})+\mathbf{1}}{2} = F((\mathbf{m}^{(\beta)}+\mathbf{\overline{e}}_1), \mathbf{h}_1) \mathbf{G_{\Lambda}}\mathbf{P}_1+\mathbf{e}_1\mathbf{P}_1.
\end{align*}
Indeed, the attacker computes
\begin{align*}
   \mathbf{v}^{(i)}= F((\mathbf{m}^{(i)}+\mathbf{\overline{e}}_1), \mathbf{h}_1) \mathbf{G_{\Lambda}}\mathbf{P}_1+\mathbf{e}_1\mathbf{P}_1,
\end{align*}
for $i=1, \cdots, 2^l$, to obtain
\begin{align*}
\sum\nolimits_{\beta= 1}^{{2^l}} {\mathbf{v}^{(\beta)}}&=\sum\nolimits_{\beta= 1}^{{2^l}} {F((\mathbf{m}^{(\beta)}+\mathbf{\overline{e}}_1), \mathbf{h}_1) \mathbf{G_{\Lambda}}\mathbf{P}_1+\mathbf{e}_1\mathbf{P}_1}\\
&=\sum\nolimits_{\beta= 1}^{{2^l}} {F((\mathbf{m}^{(\beta)}+\mathbf{\overline{e}}_1), \mathbf{h}_1) \mathbf{G_{\Lambda}}\mathbf{P}_1}+ \sum\nolimits_{\beta= 1}^{{2^l}} {\mathbf{e}_1\mathbf{P}_1}.
\end{align*}
On the other hand, according to Eq. (\ref{generator matrix}), we have
\begin{align*}
F((\mathbf{m}^{(\beta)}+\mathbf{\overline{e}}_1), \mathbf{h}_1)& \mathbf{G_{\Lambda}}\mathbf{P}_1
 =  F((\mathbf{m}^{(\beta)}+\mathbf{\overline{e}}_1), \mathbf{h}_1)_{[1:k]}\mathbf{G_{\mathcal{C}}}\mathbf{P}_1  
                  +F((\mathbf{m}^{(\beta)}+\mathbf{\overline{e}}_1), \mathbf{h}_1)_{[k+1:n]} \left[\begin{array}{cc}
                  \mathbf{0} & 2\mathbf{I_{n-k}}
                  \end{array}\right]\mathbf{P}_1,
 \end{align*}
where $\mathbf{G_{\mathcal{C}}}$ is the systematic form of the generator matrix of the used QC-LDPC code (Eq. (\ref{G-qc})) and
 $\mathbf{x}_{[1:k]}$ denotes the entries  $1$ to $k$ of the vector $\mathbf{x}$.
Therefore, recovering $\mathbf{G_{\mathcal{C}}}$ is sufficient for determining the  $\mathbf{G_{\Lambda}}$ and the attacker has the following steps to recover $\mathbf{G_{\Lambda}}\mathbf{P}_1$
\begin{align}
\sum\nolimits_{\beta= 1}^{{2^l}} {\mathbf{v}^{(\beta)}} \nonumber
&\equiv\left(\sum\nolimits_{\beta= 1}^{{2^l}} {F'((\mathbf{m}^{(\beta)}+\mathbf{\overline{e}}_1), \mathbf{h}_1)} \right) \mathbf{G_{\mathcal{C}}}\mathbf{P}_1 ~ (\textrm{mod} ~ 2),
\end{align}
where $F'(\mathbf{a}, \mathbf{b})=\{f_1(\mathbf{a}, \mathbf{b})~(\textrm{mod} ~2), \ldots, f_k(\mathbf{a}, \mathbf{b})~(\textrm{mod} ~2)\}$ and $f_i(\mathbf{a}, \mathbf{b})~ (\textrm{mod} ~2)$'s are the Boolean functions defined in Eq. (\ref{function-F}).
The algebraic degree of each component function and their nonzero linear combination is $d+1$ \cite{Function-f}. Therefore, the  algebraic degree of function $F'$ is equal to $d+1$.

Let $L[\mathbf{\mathfrak{u}}_{1}, \ldots, \mathbf{\mathfrak{u}}_{l}]$ be the list of
all $2^l$ possible linear combinations of $\mathbf{\mathfrak{u}}_{1}, \ldots, \mathbf{\mathfrak{u}}_{l}$ where  $\mathbf{\mathfrak{u}}_{i}\in \mathbb{F}_2^n$ is a vector with a $1$ in the $i$th position and zeros in all the other positions.
We recall that the $l$th  derivative of a Boolean function $f(\mathbf{x})$ is defined as $\Delta^{(l)}_{\mathbf{\mathfrak{u}}_{i_1}, \ldots,\mathbf{\mathfrak{u}}_{i_l}} f(\mathbf{x})=
\sum\nolimits_{c \in L[\mathbf{\mathfrak{u}}_{1}, \ldots, \mathbf{\mathfrak{u}}_{l}]} {f(\mathbf{x} + \mathbf{c})} $ \cite{Higher-order-derivatives}.
Therefore, since $ \mathbf{m}^{(\beta)}$s are binary vectors with zero components in the last $(n-k)$ positions,
\small
\begin{align}
\sum\nolimits_{\beta= 1}^{{2^l}} {\mathbf{v}^{(\beta)}} ~ (\textrm{mod} ~ 2)\nonumber
&=\left(\sum\limits_{\mathbf{m}^{(\beta)}\in L[\mathbf{\mathfrak{u}}_{i_1}, \ldots, \mathbf{\mathfrak{u}}_{i_l}]} {F'((\mathbf{m}^{(\beta)}+\mathbf{\overline{e}}_1), \mathbf{h}_1)} \right) \mathbf{G_{\mathcal{C}}}\mathbf{P}_1\\  \label{differential-attack}
&=\left( \Delta^{(l)}_{\mathbf{\mathfrak{u}}_{i_1}, \ldots, \mathbf{\mathfrak{u}}_{i_l}} F'(\mathbf{\overline{e}}_1, \mathbf{h}_1) \right) \mathbf{G_{\mathcal{C}}}\mathbf{P}_1.
\end{align}
\normalsize

When $l = deg(F')$ and  $\mathbf{\mathfrak{u}}_{i_1}, \ldots, \mathbf{\mathfrak{u}}_{i_l}$ are chosen in which, for $\rho=1, \ldots, l$, $i_\rho\leq k$,  then $\Delta^{(l)}_{\mathbf{\mathfrak{u}}_{i_1}, \ldots, \mathbf{\mathfrak{u}}_{i_l}} F'(\mathbf{\overline{e}}_1, \mathbf{h}_1)$ is a constant vector independent of $\mathbf{\overline{e}}_1$.
Hence, Eq. (\ref{differential-attack}) returns a linear equation with respect to $\mathbf{G_{\mathcal{C}}}\mathbf{P}_1$. If the attacker has $k$ such  linear independent equations, then he can find $\mathbf{G_{\mathcal{C}}}\mathbf{P}_1$. Therefore, the attacker computes Eq. (\ref{differential-attack}) for other messages, at least $k$ times, to prepare a system of $k$ linear equations with respect to $\mathbf{G_{\mathcal{C}}}\mathbf{P}_1$ and then solve it. The complexity of the first step of the attack is $O(k \times 2^{deg(F')})$. Since the vector $\mathbf{h}$ is secret, the attacher needs $2^{d-1}$ guesses.
Hence, the overall complexity of recovering $\mathbf{G_{\mathcal{C}}}\mathbf{P}_1$ is of order $O(k \times 2^{d-1} \times 2^{deg(F')})=O(k \times 2^{d-1} \times 2^{d+1})=O(k \times 2^{2d})$. For the proposed QC-LDPC lattice with  $(k, n)=(215, 258)$ and $d=61$, the complexity of the attack is  $2^{129}$.

The next steps of this attack  involve recovering $\mathbf{e}_1$ and decrypting the ciphertext using $\mathbf{P}_1$.
The first instance of the ciphertext is given as
\begin{eqnarray}\label{attack-on-s1-m}
 \mathbf{v}^{(1)} = F((\mathbf{m}^{(1)}+\mathbf{\overline{e}}_1), \mathbf{h}_1)\mathbf{G_{\Lambda}}\mathbf{P}_1+\mathbf{e}_1\mathbf{P}_1.
 \end{eqnarray}
The attacker applies it to $\mathbf{m}^{(1)}=(0, \ldots, 0)$, obtaining
\begin{eqnarray}\label{attack-on-s1}
 \mathbf{v}^{(1)}
                  &\equiv& F'(\mathbf{\overline{e}}_1, \mathbf{h}_1)\mathbf{G_{\mathcal{C}}}\mathbf{P}_1
                  +\mathbf{e}_1 \mathbf{P}_1 ~(\textrm{mod} ~2).
\end{eqnarray}
Since $\mathbf{G_{\mathcal{C}}}\mathbf{P}_1 $ has been estimated in the previous step, the attacker should estimate $F'(\mathbf{\overline{e}}_1, \mathbf{h}_1)$ and $F'(\mathbf{\overline{e}}_1, \mathbf{h}_1)\mathbf{G_{\mathcal{C}}}\mathbf{P}_1$ for all possible vectors $\mathbf{e}_1$ and $\mathbf{h}_1$. The corresponding $\mathbf{e}_1 \mathbf{P}_1$ can be computed using Eq. (\ref{attack-on-s1}), that is,  $\mathbf{v}^{(1)} ~(\textrm{mod} ~2)\oplus F'(\mathbf{\overline{e}}_1, \mathbf{h}_1)\mathbf{G_{\mathcal{C}}}\mathbf{P}_1=\mathbf{e}_1 \mathbf{P}_1$. Recovering the matrix $\mathbf{P}_1$, based on this equation, requires  $vq^2$ search for the different choices for $(\mathbf{e}_1, \mathbf{h}_1)$.
Indeed, Eq. (\ref{attack-on-s1})  constructs a  linear system of equations $\mathbf{x}=\mathbf{e}_1 \mathbf{P}_1$ in terms of the $\mathbf{P}_1$ entries as its variables. With $\mathbf{P}_1$ as a block diagonal matrix, we have $\mathbf{x}= ({\mathbf{e}_1}_{[1:q]}\mathbf{\pi}_{1}, {\mathbf{e}_1}_{[q+1:2q]}\mathbf{\pi}_{2}, \ldots, {\mathbf{e}_1}_{[(v-1)q+1:vq]}\mathbf{\pi}_{v})$, where ${\mathbf{e}_1}_{[(i-1)q+1:iq]}$ is the $i$th $q$ bits of $\mathbf{e}_1$.
Moreover, each sub-permutation matrix has one nonzero component in each column, leading to the equation $\mathbf{x}_j={\mathbf{e}_1}_{[(j-1)q+1:jq]}\mathbf{\pi}_{j}$ has $q^2$ solutions. Therefore, there are $vq^2$ solutions for $\mathbf{P}_1$ in this system. The complexity of finding $\{\widetilde{\mathbf{e}_1}, \widetilde{\mathbf{h}_1}, \widetilde{\mathbf{P}_1}\}$ that satisfy  Eq. (\ref{attack-on-s1}) is  $O(2^{(l_1+l_2)}\times vq^2)$.
The attacker repeats the same work to find $\{\overline{\mathbf{e}_1}, \overline{\mathbf{h}_1}, \overline{\mathbf{P}_1}\}$  satisfying Eq. (\ref{attack-on-s1-m}) using an arbitrary message  with complexity  $O(2^{(l_1+l_2)}\times vq^2)$. Then, the attacker compares  $\{\widetilde{\mathbf{e}_1}, \widetilde{\mathbf{h}_1}, \widetilde{\mathbf{P}_1}\}$ and  $\{\overline{\mathbf{e}_1}, \overline{\mathbf{h}_1}, \overline{\mathbf{P}_1}\}$ in order to verify the suitable set for this step.
Hence, the complexity of estimating   $\{\mathbf{e}_1, \mathbf{h}_1, \mathbf{P}_1\}$ is approximately $O(2^{(l_1+l_2+1)}\times 2vq^2)$, which is almost $O(2^{84})$ for the proposed scheme.
At this moment, the  attacker  can decrypt any $\mathbf{v}^{(1)}$ using $\{\mathbf{e}_1, \mathbf{h}_1, \mathbf{P}_1\}$ and $\mathbf{G_{\mathcal{C}}}\mathbf{P}_1$. Hence, the overall complexity for decrypting $\mathbf{v}^{(1)}$  is $O(k \times 2^{2d}+2^{(l_1+l_2+1+1)}\times 2vq^2)\approx O(2^{129})$.
Since the intentional error vector and permutation matrix change with each message, the encryption matrix $\mathbf{G_{\mathcal{C}}}\mathbf{P}_i$ and each vector $\mathbf{e}_i\mathbf{P}_i$   change in the $i$th round of encryption, while $\mathbf{h}_i$s are known after estimation of $\mathbf{h}_1$.  Therefore, for recovering all keys, the attacker can consider the zero vector as a message and compute the ciphertext $\mathbf{v}^{(2)} = F(\mathbf{\overline{e}}_2, \mathbf{h}_2)\mathbf{G_{\Lambda}}\mathbf{P}_2+\mathbf{e}_2\mathbf{P}_2$.

Next, $\mathbf{G_{\mathcal{C}}}\mathbf{P}_2$ is estimated with complexity of order $O(k \times 2^{deg(F')})=O(k \times  2^{d+1})$, approximately $O(2^{70})$ in our case. Furthermore, $\mathbf{P}_2$ is computed based on the above equation of $\mathbf{v}^{(2)}$ in the similar way as mentioned above. The complexity of computing the other encryption matrix $\mathbf{G_{\mathcal{C}}}\mathbf{P}_i$, for $i=3, \ldots $, in each round is $O(2^{70})$. Thus, the overall complexity of  this attack is $O( 2^{129}+N_p\times 2^{70})$, where $N_p$ is the number of possible permutations. As a consequence, the overall complexity for the proposed parameters is $O( 2^{129}+2^{36}\times 2^{70}) \approx O( 2^{129})$. Therefore, with the proposed parameters, we achieve a security level of $128$-bit.

\section{Comparison}~\label{sec:Comparison}

In this section, we compare our proposed lattice based cryptosystem with \cite{Hooshmand},  \cite{Deepthi-Nonlinear}  and\cite{GLOBECOM}, in which
 \cite{Hooshmand} is a lattice based joint scheme (based on  LDLC lattices) and the others are based on error correcting codes.
Moreover, in \tablename~\ref{comparison}, we review  a number of previous schemes  that join error correction and encryption in one process to enable efficient implementations.
Although, the existing schemes  based on lattices and   error correcting codes are not in the same category,
 we also point out the results of the joint schemes based on error correcting codes to provide an intuition about the counterpart of lattice based schemes.

The recent work of Stuart and Deepthi \cite{Deepthi-Nonlinear} is an RN-like scheme based on QC-LDPC codes that strengthen the cryptosystem against differential attacks.  It has the smallest key size among other RN-like schemes and its complexity of encryption and decryption is of  $O(n^2)$ due to the used nonlinear function in its design.
Similar to other RN-like schemes, the output of this scheme is fed into  a modulator to transmit through an AWGN channel.
\begin{table*}[!ht]
\scriptsize
\caption{Comparison of the proposed scheme with other secure channel coding schemes.}
\centering
\begin{tabular}{|c||c|c|c|c|}
\hline
Cryptosystem   & Underlying code, $\mathcal{C}(n,k)$ & Information  rate  & key size \\
  \hline
   \hline
Rao \cite{Rao}    & Goppa code, $\mathcal{C}(1024, 524)$, coding rate$=0.51$ & N/A &  $2$ Mbits  \\
  \hline
Rao and Nam  \cite{Rao-Nam}   & Hamming code, $\mathcal{C}(72, 64)$, coding rate$=0.89$ & N/A &  $18$ kbits  \\
  \hline
Sobhi Afshar et al. \cite{Eghlidos} &   EG-QC-LDPC code, $\mathcal{C}(2044, 1024)$, coding rate$=0.5$ & N/A &  $2.5$ kbits  \\
  \hline
Hooshmand et al. \cite{Hooshmand-ISC}   & EDF-QC-LDPC code, $\mathcal{C}(2470, 2223)$, coding rate$=0.9$ & N/A &  $3.55$ kbits  \\
  \hline
Esmaeiliet et al. \cite{Esmaeili1}    & QC-LDPC code,  $\mathcal{C}(2048, 1536)$, coding rate$=0.75$ & N/A &   $2.191$ kbits \\
  \hline
Esmaeili and Gulliver \cite{Esmaeili2}    & QC-LDPC code,  $\mathcal{C}(2048, 1536)$, coding rate$=0.75$ & N/A &   $2.22$ kbits \\
  \hline
   Adamo et al. (ECBC) \cite{ECBC} &  LDPC code, $\mathcal{C}(256, 128)$, coding rate$=0.5$ & N/A & 82 kbits.\\
  \hline
Pisek  et al. \cite{GLOBECOM}   & QC-LDPC code, $\mathcal{C}(128, 256)$, coding rate$=0.5$ & N/A & $128$ bits\\
  \hline
Stuart and Deepthi \cite{Deepthi-Nonlinear}  & EDF-QC-LDPC code, $\mathcal{C}(124, 248)$, coding rate$=0.5$ & N/A & $182$ bits \\
 \hline
LDLC lattice based cryptosystem \cite{Hooshmand}   & Latin square LDLC lattice, $n=10^4$ symbols & N/A &  $3$ Mbits  \\
  \hline
The proposed lattice based cryptosystem &   RDF-QC-LDPC code, $\mathcal{C}(258, 215)$, $L=16$ &  $5$ &  $214$ bits  \\
  \hline
 \end{tabular}
 \label{comparison}
 \end{table*}
 \normalsize

The most efficient joint AES-coding scheme is proposed in  \cite{GLOBECOM}, where
QC-LDPC  codes are embedded in  each  round of the AES encryption and decryption. This scheme consists of two parts for encryption and channel coding, with the encryption part which is more powerful than classical AES. The coding part of the proposed scheme outperforms other conventional joint AES-coding schemes. It applies the same parity-check matrix for the encryption and encoding parts. Promising ideas such as using  lower triangular matrices and the quasi-cyclic structure of the LDPC code and using the same hardware resource for both parts, reduce  power consumption compared to other joint AES-coding schemes \cite{GLOBECOM}. Encryption and channel coding cannot be applied simultaneously in \cite{GLOBECOM} and the output of encryption is fed into the encoder and then the resulting data is passed through the QPSK modulator to transmit via an AWGN channel.

Unlike the proposed schemes in  \tablename~\ref{comparison}, we have merged these three steps into a single step using efficient lattices  in our cryptosystem.
This provides less delay, lower implementation complexity and better  error performance in overall for high SNR or  bandlimited channels.
The column three in \tablename~\ref{comparison} describes the average information bits per symbol in the transmission which is not available for other schemes. If  we want compute  the average information bits per symbol in the transmission of a joint scheme, the coding rate of the used code in the structure of the scheme should multiply with the rate of a modulation that will be applied after.
Since after encoding the rate of the joint schemes is less than  $1$, they should use high order modulation to compensate this low rate for bandlimited AWGN channels.
For example, with the proposed parameters in  \tablename~\ref{comparison}, the information rate of our scheme is $5$. To reach the same information rate with the code based schemes like \cite{Deepthi-Nonlinear} and \cite{GLOBECOM}, which have coding rate $0.5$, should use a high order modulation like $1024-$QAM  after coding step ($0.5\times 10=5$). However, applying high order modulation makes the error performance weaker.
Therefore, code based schemes are not appropriate for bandlimited channels without combining them with high order modulation.  In contrast, the lattice codes are high order modulation schemes with error correction capability and provide high information rate.
Since there are not such a complete setting  about concatenation of AES-joint schemes or RN-like schemes and a high order modulation in the literature, we can not have a fair comparison about  the  overall error performance  of joint schemes based on codes and our proposed scheme. Our lattice based scheme is the first candidate introducing a scheme for secure communication on bandlimited channels.

In the sequel, we compare the key size of our cryptosystem with the LDLC lattice based scheme. In  \cite{Hooshmand}, Latin square LDLCs are used to provide a joint encryption and encoding scheme.
The  encryption and decryption complexity of the proposed LDLC  based scheme is $O(n^2\beta)$ and $O(n^2\delta)$, respectively, where
$\beta$ and  $\delta$ are the maximum required memory to save each entry of its rational generator matrix and its secret key in  binary form, respectively, and  $n$ is the lattice dimension \cite{Hooshmand}.
On the other hand, the ciphertext $\textbf{c}$ is transmitted through an ``unconstrained'' power AWGN channel.
Indeed, the shaping is removed to decrease the encryption complexity, while the computational complexity of the encryption/encoding grows rapidly due to the existence of unbounded lattice points and its average transmitted power becoming too large.

The Latin square LDLC lattices are introduced by the parity-check matrix $\textbf{H}$ which is an $n\times n$  Latin square. This matrix is determined by a generating sequence set $\mathcal{H}=\{h_1, h_2, \ldots, h_d\}$, where $h_i$s are nonzero values at the appropriate locations of the used Latin square $\textbf{H}$. The generating sequence set $\mathcal{H}$ is considered as the secret key  of the LDLC  based scheme, instead of its corresponding parity-check matrix $\textbf{H}$ \cite{Hooshmand}.
In this way, the legitimate receiver  has to construct the same parity-check matrix $\textbf{H}$ using the secret key $\mathcal{H}$ and is able to recover the original message in decryption. It can be shown that the generating set $\mathcal{H}$ does not result in a unique parity-check matrix for the Latin square LDLC that is used for decryption, and then the qualified receiver is not able to decrypt the ciphertexts correctly.
As there are $\mathcal{L}(n)=n! \sum\nolimits_{\textbf{A}\in B_n}{(-1)^{\sigma_0(\textbf{A})}}{per(\textbf{A})\choose n}$,
$n \times n$ Latin squares, where $ B_n$ is the set of all binary $n \times n$  matrices, $\sigma_0(\textbf{A})$ is the number of zero entries in matrix $\textbf{A}$ and $per(\textbf{A})$ is the permanent of matrix $\textbf{A}$ \cite{Latin-squares}, there are at least $\mathcal{L}(n)/(n-d)!$
different parity-check matrices for Latin square LDLCs with the given generating set  $\mathcal{H}$.
Thus, for successful  decryption, the generating set $\mathcal{H}$ should be replaced with the parity-check matrix $\textbf{H}$ as the secret key in \cite{Hooshmand}.  In this way, since the position of each non-zero entry along with its value has to be saved, the memory consumption for saving $\textbf{H}$  is $ nd\big(r+2\lceil\log_2\left( n\right)\rceil\big) $ bits, , where $r$ is the maximum number of bits required for saving $h_i$, for $i=1,\ldots, d$.

The secret key of the proposed  LDLC based scheme is $\mathcal{K}=\{\mathcal{H},\mathcal{ P}\}$, where $\mathcal{P} =\{p_1, p_2, \ldots, p_d\}$ is the set of indexes   such that $1\leq p_i\leq n$, for $1\leq i\leq d$ \cite{Hooshmand}. Therefore, the key size of an LDLC based cryptosystem, when considering the parity-check matrix $\textbf{H}$, is at most $d \big( n(r+2\lceil\log_2\left( n\right)\rceil)+ \lceil\log_2\left( n\right)\rceil \big)$ bits.
In the same level of security (128 bits), the key size of the LDLC based cryptosystem for an LDLC with parameters $ n=10^4, d=7$ and  $r=16$ is $3080098$ bits; while the key size for our cryptosystem is equal to $214$ bits with a   $(258, 215)$-QC-LDPC code, $b=q=43, d_v=3, n_0=6$ and $d_c=18$  that is much smaller.  As a consequence, our proposed cryptosystem has a small key size compared to other RN-like cryptosystems that can also do modulation simultaneously.
\section{Conclusion}~\label{sec:Conclusion}
In this paper, we have proposed a new RN-like encryption scheme using  QC-LDPC lattices.
Moreover, we have exploited lattice codes related to these lattices to join encryption, channel coding and modulation in a single step that is suitable for resource limited applications. The proposed nonlinear cryptosystem is secure against all variants of chosen plaintext attacks against RN-like encryption schemes.
The main advantages of the proposed scheme are its high information rate, small key size and low hardware complexity.
As a consequence, the joint scheme  provides high speed and efficient implementation as well as secure and reliable data transmission for bandlimited AWGN channels.
\ifCLASSOPTIONcaptionsoff
  \newpage
\fi

\end{document}